\documentclass[a4paper,twocolumn,english,aps,prl,superscriptaddress]{revtex4-1}
\usepackage[T1]{fontenc}
\usepackage[latin9]{inputenc}
\usepackage{color}
\usepackage{babel}
\usepackage{bm}
\usepackage{amsmath}
\usepackage{amssymb}
\usepackage{graphicx}
\usepackage[unicode=true,pdfusetitle,
 bookmarks=true,bookmarksnumbered=false,bookmarksopen=false,
 breaklinks=false,pdfborder={0 0 1},backref=false,colorlinks=false]
 {hyperref}
\usepackage{breakurl}

\makeatletter

\usepackage{babel}
\usepackage{babel}

\makeatother

\begin{document}
\renewcommand{\figurename}{Fig.}
\title{Enhancement of Second-Order Non-Hermitian Skin Effect by Magnetic
Fields}
\author{Chang-An Li}
\affiliation{Institute for Theoretical Physics and Astrophysics, University of
Würzburg, 97074 Würzburg, Germany}
\author{Björn Trauzettel}
\affiliation{Institute for Theoretical Physics and Astrophysics, University of
Würzburg, 97074 Würzburg, Germany}
\author{Titus Neupert}
\affiliation{Department of Physics, University of Zürich, Winterthurerstrasse 190,
8057, Zürich, Switzerland}
\author{Song-Bo Zhang}
\email{songbo.zhang@physik.uzh.ch}

\affiliation{International Center for Quantum Design of Functional Materials (ICQD), Hefei National Research Center for Physical Sciences at the Microscale, University of Science and Technology of China, Hefei, Anhui 230026, China}
\affiliation{Department of Physics, University of Zürich, Winterthurerstrasse 190,
8057, Zürich, Switzerland}
\date{\today}
\begin{abstract}
The non-Hermitian skin effect is a unique phenomenon in which an extensive
number of eigenstates are localized at the boundaries of a non-Hermitian
system. Recent studies show that the non-Hermitian skin effect is
significantly suppressed by magnetic fields. In contrast, we demonstrate that the second-order skin effect (SOSE) is robust
and can even be enhanced by magnetic fields. Remarkably, SOSE can
also be induced by magnetic fields from a trivial non-Hermitian system
that does not experience any skin effect at zero field. These properties
are intimately related to to the persistence and emergence of topological
line gaps in the complex energy spectrum in presence of magnetic fields. Moreover, we show that a magnetic field can drive
a non-Hermitian system from a hybrid skin effect, where the first-order
skin effect and SOSE coexist, to pure SOSE. Our results describe
a qualitatively new magnetic field behavior of the non-Hermitian skin
effect.
\end{abstract}
\maketitle
\textit{\textcolor{blue}{Introduction.}}---Non-Hermitian physics appears in a variety
of physical systems in quantum optics, ultracold atoms, and condensed
matter\ \cite{Bender07RPP,El-Ganainy18nphys,Ashida20AP,Bergholtz21rmp}.
In recent years, many unique non-Hermitian phenomena have been revealed,
which show no counterparts in their Hermitian limit\ \cite{Lee16prl,Yao18prl,Kunst18prl,ZhangK20prl,Okuma20prl,ZhangX22AP,Okuma23AR,Yao18prl2,GonZP18prx,ShenH18prl,Esaki11prb,Leykam17prl,Lieu18prb,XiongY18jpc,Yin18prb,Alvarez18prb,Kawabata19PRL,Yokomizo19prl,Zhou19prb,LeeJY19prl,RuiWB19PRB,Okugawa19prb2,Budich20prl,HuHP21,Wojcik20prb,Kornich22prr,ZhangSB22prb}.
One prime example is the non-Hermitian skin effect (NHSE), an anomalous
localization of extensive eigenstates at the open boundaries of a
non-Hermitian system\ \cite{Lee16prl,Yao18prl,Kunst18prl,ZhangK20prl,Okuma20prl,ZhangX22AP,Okuma23AR}.
This intriguing phenomenon has triggered intense research interests
both theoretically and experimentally\ \cite{Lee16prl,Yao18prl,Kunst18prl,ZhangX22AP,Okuma23AR,Yao18prl2,Kawabata19prx,
Ezawa19prb2,LiuT19prl,LeeCH19prb,Longhi19prl,XWLuo19prl,Imura19prb,Jin19prb,Herviou19pra,ZhangK20prl,Okuma20prl,ZhangDW20scp,Xiao20np,LiLH20prl,
Hofmann20prr,Matsumoto20prl,LiLH20nc,GaoP20prl,Ghatak20pnas,YangZS20prl,Helbig20np,Suthar22PRB,
Yokomizo20prr,Denner21nc,Okuma21prb,Zirnstein21prl,Haga21prl,Claes21prb,GuoC21prl,ParkMJ21prb,
Schindler21prb,Yokomizo21prb,SunXQ21prl,WuD22prb,Zhangk22nc,Alsallom22prr,Franca22prl,QinF23pra,
XueW22prl,Jezequel23prl,Wang22arxiv,Okugawa23arxiv}.
In the conventional (first-order) NHSE, the skin modes are localized
at $(d-1)$ or lower dimensional boundaries, and importantly, their
number scales with the volume $L^{d}$, where $d$ and $L$ are dimension
and linear size of the system, respectively~\cite{Kawabata20prb}.
Recently, NHSE has been generalized to higher ($n$-th) orders with
certain constraints on the geometry, where the skin modes are localized
to $(d-n)$ or lower dimensional boundaries and their number scales
as $L^{d-n}$ with $2\leqslant n\leqslant d$\ \cite{LeeCH19prl,Kawabata20prb,Okugawa20prb}.
Notably, SOSE (with $n=2$) has been predicted in several lattice
models and implemented in several experimental platforms\ \cite{LeeCH19prl,Kawabata20prb,Okugawa20prb,Okugawa21prb,FuY21prb,LiY22prl,Zhangx21nc,Palacios21nc,Zou21nc,Zhuw22prb,ShangC22advs}.

In presence of uniform magnetic fields, charged particles display
localization behavior due to cyclotron motion in the plane perpendicular
to the field\ \cite{Klitzing80prl,cage2012quantum}. Hence, the magnetic
field tends to localize most particles in the bulk of the system in
contrast to NHSE. It is thus interesting to investigate the coexistence
of NHSE and magnetic fields. So far, research efforts have been mostly
dedicated to the first-order skin effect (FOSE) with the outcome that
it is significantly suppressed even by small magnetic fields due to
competing tendencies of localization\ \cite{Okuma21prl,LuM21prl,ShaoK22prb}.
However, as we show below, this physical picture does not apply to
the higher-order NHSE.

\begin{figure}
\includegraphics[width=1\linewidth]{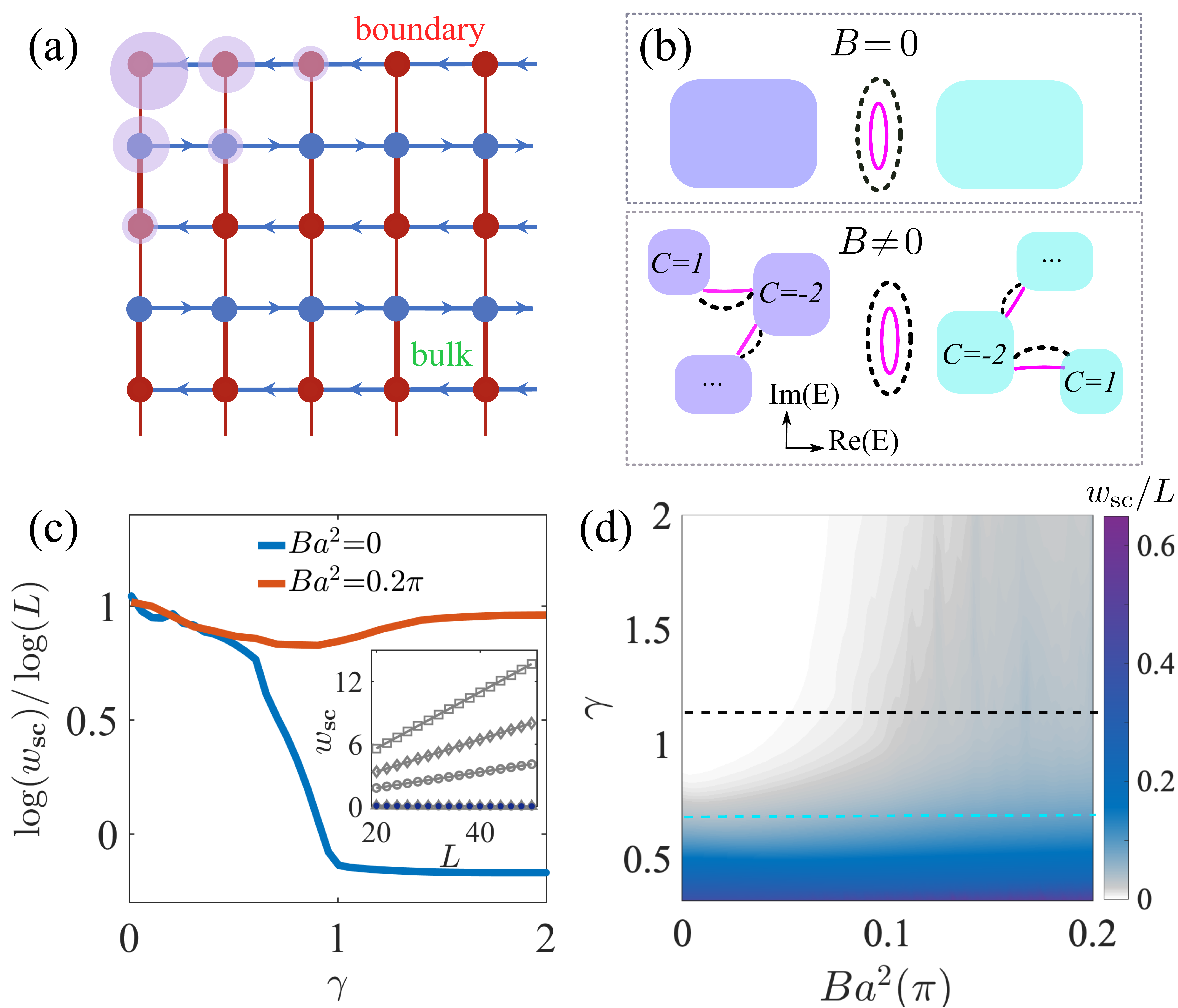}

\caption{(a) Schematic of the model in Eq.~\eqref{eq:Hamiltonian1} with OBC
in $x$- and $y$-directions. The red thick and thin bonds indicate
dimerized hoppings in $y$-direction; the blue arrow lines indicate
non-reciprocal hoppings in $x$-direction, and the purple dots mark
the SCMs. (b) Sketch of the energy spectrum (colored boxes) in absence
(top) and presence (bottom) of a magnetic field. Solid (dashed) lines
indicate the spectrum of SCMs (edge modes). (c) $\log(w_{\text{sc}})/\log(L)$
at $Ba^{2}=0$ (blue) and $0.2\pi$ (orange) as a function of $\gamma$
for $L=160$ and $\xi=30$, respectively. Inset: skin corner weight
$w_{\text{sc}}$ as a function of $L$ at $B=0$ for $\xi=1$, $\gamma=0.4$,
$0.5$, $0.6,$ $1.0$ and $1.6$ (from top to bottom). (d) $w_{\text{sc}}/L$
as a function of $\gamma$ and $B$ for $L=60$. $\lambda=1$ for
(c,d). \label{fig1:main-result}}
\end{figure}

In this work, we study systematically the influence of magnetic fields
on SOSE in two dimensions (2D). Remarkably, we find that SOSE is not
suppressed but rather enhanced by magnetic fields. It can even be
induced by magnetic fields in otherwise trivial non-Hermitian systems
without any skin effect. We explain such anomalous behaviors by topological
properties of the complex energy spectrum. We demonstrate
the results on several prototypical non-Hermitian models for SOSE
with line-gap or point-gap topology. In the latter case, we further
show a magnetic field-driven transition from a hybrid skin effect
with coexistence of FOSE and SOSE to pure SOSE. Finally, we discuss
feasible experimental platforms to test our predictions.

\textit{\textcolor{blue}{Phenomenology of SOSE and its magnetic robustness.}}---To
better illustrate our results, we first consider 2D non-Hermitian
systems in which FOSE is absent but SOSE can be present. The absence
of FOSE is achieved by imposing certain spatial symmetries to the
bulk of the system, such as mirror and inversion symmetries, together
with an appropriate choice of open boundary conditions (OBC)\ \cite{Li2022footnote1-FOSE}.
SOSE in the non-Hermitian system usually arises as a consequence of
the interplay between 1D edge modes and local non-reciprocity\ \cite{LeeCH19prl,Kawabata20prb}.
Note that the spatial symmetries are locally violated at the boundary.
If the system has edge modes along the boundary with OBC in one direction
and periodic boundary conditions (PBC) in the other direction, these
edge modes will experience non-reciprocity. Thus, upon further imposing
OBC to the other direction, they evolve into skin corner modes (SCMs)
{[}Fig.\ \ref{fig1:main-result}(a){]}. The number of SCMs scales
with the linear system size $\text{\ensuremath{\sim}}\mathcal{O}(L)$,
as inherited from the edge modes. This is the hallmark of SOSE\ \cite{Kawabata20prb}.

The properties of SOSE in this case depend substantially on the edge
modes, which can be of topological origin. Similar to the Hermitian
case, the edge modes are typically protected by bulk gaps, referred
to as line gaps in the non-Hermitian context\ \cite{Kawabata19prx,ShenH18prl,Borgnia20prl}. Indeed, we observe a coincidence between the appearance/disappearance
of SOSE and line-gap transitions. The line-gap topology persists under
the application of magnetic fields {[}Fig.\ \ref{fig1:main-result}(b){]}.
Thus, the robustness of SOSE based on edge modes relates to a line-gap protection. Furthermore, under strong magnetic fields, each bulk
continuum spectral area generally splits into multiple non-Hermitian
Hofstadter bands {[}lower panel in Fig.\ \ref{fig1:main-result}(b){]}.
Chiral edge modes appear to connect the Hofstadter bands with different
Chern numbers when OBC are applied in one direction. Due to the local
non-reciprocity at the boundary, these emerging edge modes likewise
evolve into SCMs when OBC are further imposed in the other direction.
Thus, the system exhibits enhanced SOSE due to magnetic fields.

In non-Hermitian systems with point-gap topology,
we can observe the occurrence of a hybrid skin effect, where FOSE
and SOSE coexist. Applying a magnetic field significantly suppresses
FOSE, while it enhances SOSE\ \cite{Li2022footnote4-linegap}. As
a result, the system transforms from the hybrid skin effect at zero
field to pure SOSE state at strong fields. We illustrate these intriguing
behaviors within explicit models below {[}cf. Eqs.~\eqref{eq:Hamiltonian1}
and \eqref{eq:Hamiltonian2}{]}.

\textit{\textcolor{blue}{Skin corner weight.}}---To quantitatively
characterize SOSE under OBC, we introduce the skin corner weight $w_{\text{sc}}$
defined as
\begin{equation}
w_{\text{sc}}\equiv\sum_{n,{\bf r},{\bf r}_{\text{c}}}\left|\psi_{n}^{R}({\bf r})\right|^{4}\exp(-\left|{\bf r}-{\bf r}_{\text{c}}\right|/\xi),\label{eq:skinw}
\end{equation}
where the sums run over all eigenstates indexed by $n$, lattice sites
${\bf r}$, and corner positions ${\bf r}_{\text{c}}$. $\psi_{n}^{R}({\bf r})$ is the right eigen wavefunction of the Hamiltonian\ \cite{Li2022footnote2-LeftEigen}.
$w_{\text{sc}}$ in Eq.\ \eqref{eq:skinw} contains two important
factors. The first factor $\left|\psi_{n}^{R}({\bf r})\right|^{4}$
measures the localization strength of eigenstates. With only this
factor, $w_{\text{sc}}$ reduce\textcolor{black}{s} to the inverse
participation ratio which characterizes the localization properties
of the whole system. The second factor $\exp(-\left|{\bf r}-{\bf r}_{\text{c}}\right|/\xi)$,
which decays exponentially away from the corners with decay length
$\xi$, selects only the modes localized at the corners. Note that
$\xi$ should be chosen much smaller than the linear system size,
i.e., $\xi\ll L$. For SOSE, $w_{sc}$ is finite and scales linearly
with $L$ in 2D\ \cite{Li2022footnote3-FOSE}. We can apply a power-law scaling $w_{sc}\propto L^{\alpha}$ to characterize
$w_{sc}$ and extract the scaling exponent $\alpha$. Thus, $\alpha=1$
is expected for 2D systems with pure SOSE.

\textit{\textcolor{blue}{Minimal model for SOSE.}---}To illustrate the essential physics,
we first consider a 2D minimal model for SOSE on a square lattice
with two sublattices\ \cite{Kawabata20prb}. The Hamiltonian in momentum
space reads
\begin{alignat}{1}
H({\bf k}) & =\lambda\sin k_{y}\sigma_{1}+(\gamma+\lambda\cos k_{y})\sigma_{2}\nonumber \\
 & \ \ \ +\lambda\sin k_{x}\sigma_{3}-i(\gamma+\lambda\cos k_{x}),\label{eq:Hamiltonian1}
\end{alignat}
where ${\bf k}=(k_{x},k_{y})$ is the wavevector, $\lambda$ and $\gamma$
are real parameters, and \{$\sigma_{1},\sigma_{2},\sigma_{3}$\} are
Pauli matrices in sublattice space. We set the lattice constant to
$a=1$. The model features dimerized reciprocal hopping along $y$-direction
and opposite non-reciprocal hopping for the two sublattices along
$x$-direction {[}Fig.\ \ref{fig1:main-result}(a){]}. The bulk energy
spectrum covers finite areas in the complex plane, which indicates
a geometry-dependent FOSE\ \cite{Zhangk22nc}. However, the model
respects transpose-mirror symmetries in $x$- and $y$-directions,
as indicated by $\sigma_{1}H^{T}({\bf k})\sigma_{1}=H(-k_{x},k_{y})$
and $\sigma_{3}H^{T}({\bf k})\sigma_{3}=H(k_{x},-k_{y}),$ respectively.
The spectrum for any given $k_{x}$ (or $k_{y}$) consists of two
open arcs with no interior in the complex plane\ \cite{Li2023SM}.
Thus, FOSE is absent in a rectangular geometry with OBC in $x$- and
$y$-directions, as we confirm numerically below\ \cite{Kawabata20prb,Li2023SM}.

The appearance of SOSE in the model can be understood as follows.
In absence of hopping along $x$-direction, the model decouples into
identical Su-Schrieffer-Heeger chains along $y$-direction. Under
OBC in $y$-direction, localized edge modes appear in the parameter
regime $|\gamma|<|\lambda|$. When the non-reciprocal hoppings along
$x$-direction are turned on, they cancel out effectively due to the
transpose-mirror symmetries {[}Fig.\ \ref{fig1:main-result}(a){]}.
However, the hopping along $x$-direction remains locally non-reciprocal
at the boundaries, since the transpose-mirror symmetries are locally
violated there. Thus, when OBC are imposed in both directions, the
edge modes evolve into SCMs, leading to SOSE. In the other regime
$|\gamma|\geq|\lambda|$, there are no edge modes and thus no SOSE.
These features are confirmed numerically by $w_{sc}$ in Fig.\ \ref{fig1:main-result}(c).
Clearly, $w_{\text{sc}}$ is finite and increases monotonically with
decreasing $\gamma$ for $|\gamma|<|\lambda|$, while it is negligibly
small for $|\gamma|\geq|\lambda|$ {[}Fig.\ \ref{fig1:main-result}(c),
inset{]}. Figure\ \ref{fig1:main-result}(c) illustrates $\log(w_{\text{sc}})/\log(L)$
as a function of $\gamma$\ \cite{Li2022footnote4-alpha}. We find
that $\log(w_{\text{sc}})/\log(L)$ is a negative constant for $|\gamma|\geq|\lambda|$,
whereas it saturates quickly to $1$ as $\gamma$ decreases for $|\gamma|<|\lambda|$.
Thus, we observe the expected behavior of $w_{sc}$ to properly characterize
SOSE.

\begin{figure}
\includegraphics[width=1\linewidth]{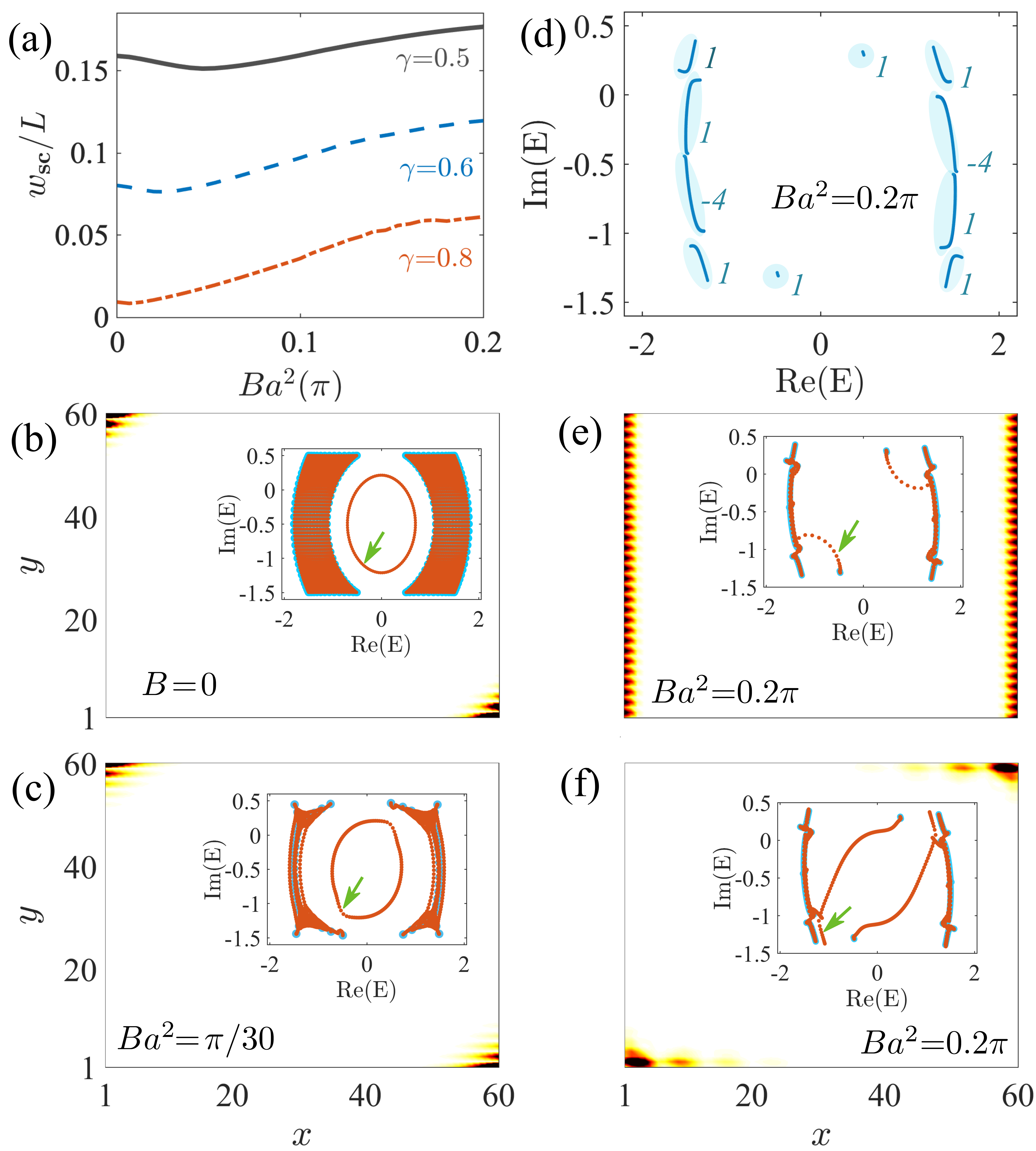}

\caption{(a) $w_{\text{sc}}/L$ as a function of $B$ for $\xi=1$ at different
$\gamma$. (b) $|\psi^{R}({\bf r})|^{2}$ of a SCM at $B=0$ under
OBC. Inset: energy spectra under PBC (cyan) and OBC (orange), respectively.
The arrow marks the mode energy $E\text{\ensuremath{=}}\text{\ensuremath{-}}0.46\text{\ensuremath{-}}1.03i$.
(c) the same as (b) but at $Ba^{2}\text{\ensuremath{=}}\pi/30$. The
arrow marks the mode energy $E=-0.53\text{\ensuremath{-}}1.10i$.
(d) Energy spectrum at $Ba^{2}=0.2\pi$ under PBC. The Chern numbers
of the Hofstadter bands are indicated. (e) $|\psi^{R}({\bf r})|^{2}$
of an edge mode at $Ba^{2}=0.2\pi$ under PBC (OBC) in $y$($x$)-direction.
The arrow marks the mode energy $E=-0.68\text{\ensuremath{-}}0.96i$.
(f) $|\psi^{R}({\bf r})|^{2}$ of a corner mode at $Ba^{2}=0.2\pi$
under OBC. Insets in (e-f): the corresponding energy spectra. The
arrow marks the mode energy $E=-1.16\text{\ensuremath{-}}1.19i$.
$\gamma=0.5$ in (b-f) and other parameters are $L=60$, $\lambda=1$.}

\label{fig2:line-gap}
\end{figure}

\textit{\textcolor{blue}{Magnetic robustness and enhancement of SOSE.}---}Now, we apply
a magnetic field $B$ to the system and study how it affects SOSE.
Without loss of generality, we adopt the Landau gauge ${\bf A}=(-By,0,0)$. 
Figures~\ref{fig1:main-result}(d) and \ref{fig2:line-gap}(a)
display $w_{\text{sc}}$ as a function of  $B$ for different $\gamma$.
Strikingly, $w_{\text{sc}}$ is not really suppressed by $B$ for
any $|\gamma|<|\lambda|$. Instead, it remains approximately a constant
for small $B$ and increases significantly as $B$ increases. This
indicates that SOSE is robust and becomes even more pronounced under
magnetic fields, in stark contrast to FOSE\ \cite{LuM21prl,ShaoK22prb,Okuma21prl}.

To better understand the line-gap protection behind the magnetic robustness
and enhancement, we analyze the energy spectrum of the system. At
zero field, the bulk spectrum consists of two detached bands separated
by a line gap {[}Fig.\ \ref{fig2:line-gap}(b), inset{]}. In a ribbon
geometry along $x$-direction, there are edge modes with energies
lying in the line gap when $|\gamma|<|\lambda|$\ \cite{Li2023SM},
forming a closed loop with point-gap topology in the complex-energy
plane. They evolve into SCMs of SOSE when further imposing OBC in
$x$-direction {[}Fig.\ \ref{fig2:line-gap}(b){]}. Note that these
features are present as long as the line gap persists in the bulk
spectrum. When gradually applying a magnetic field, the bulk spectrum
changes adiabatically, while keeping the line gap open {[}Fig.\ \ref{fig2:line-gap}(c){]}.
Therefore, SOSE stays robust against magnetic fields. Notably, the
line gap closes at $|\gamma|=|\lambda|$ and reopens for $|\gamma|>|\lambda|$,
indicating a topological phase transition. Consequently, SOSE disappears
for $|\gamma|>|\lambda|$. In this sense, the robustness of SOSE is
protected by a nontrivial line gap in the bulk spectrum.

With increasing $B$, the two bulk bands evolve into multiple non-Hermitian
Hofstadter bands\ \cite{Hofstadter76prb} that are well separated
by line gaps {[}Fig.\ \ref{fig2:line-gap}(d){]}. We find that each
Hofstadter band carries a nonzero Chern number\ \cite{Li2023SM,Hatsugai06prb}.
When OBC are applied in either direction, edge modes emerge, which
connect one Hofstadter band to another with different Chern numbers
{[}Fig.\ \ref{fig2:line-gap}(e){]}. These edge modes evolve to SCMs
when we impose OBC in both directions {[}Fig.\ \ref{fig2:line-gap}(f){]}.
Notably, the line gaps separating different Hofstadter bands remain
over a broad range of $B$, as can be seen in the corresponding Hofstadter
spectra\ \cite{Li2023SM}. This results in more pronounced SOSE
under magnetic fields.

\begin{figure}[t]
\includegraphics[width=1\linewidth]{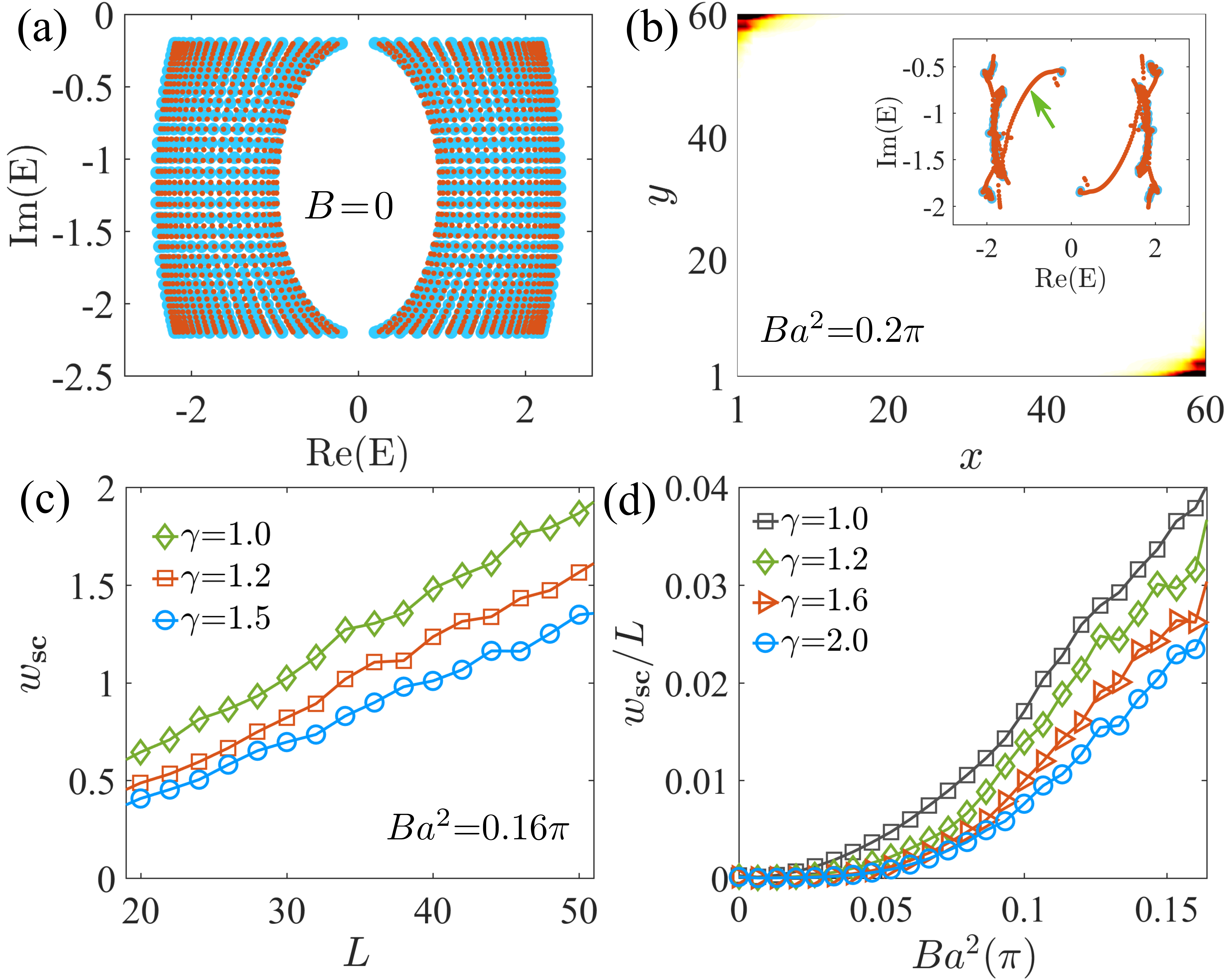}

\caption{(a) Energy spectra at $B=0$ under OBC (orange) and PBC (cyan), respectively.
(b) $|\psi^{R}({\bf r})|^{2}$ of a SCM at $Ba^{2}=0.2\pi$ under
OBC. Inset: the corresponding energy spectrum. The arrow marks the
mode energy $E=-1.00\text{\ensuremath{-}}0.73i$. (c) $w_{\text{sc}}$
as a function of $L$ at $Ba^{2}=0.16\pi$ for different $\gamma\geqslant1$.
(d) $w_{\text{sc}}/L$ as a function of $Ba^{2}$ for $\gamma\geqslant1$.
$L=60$ for (a,b,d), $\gamma=1.2$ for (a,b) and other parameters
are $\lambda=1$ and $\xi=1$.}

\label{fig3:induceSOSE}
\end{figure}

\textit{\textcolor{blue}{SOSE induced by magnetic fields.}---}We demonstrate that SOSE
can even be induced by a magnetic field in the regime $|\gamma|\geq|\lambda|$.
At $B=0$, the system has nearly the same energy spectrum under different
boundary conditions {[}Fig.\ \ref{fig3:induceSOSE}(a){]}. Accordingly,
it does not have SCMs under OBC. However, in presence of $B$, the
boundary spectrum under OBC becomes substantially different from that
under partial PBC {[}Fig.\ \ref{fig3:induceSOSE}(b), inset{]}. In
particular, under OBC, we observe eigenstates whose energies strongly
deviate from both bulk and edge spectra. In fact, these eigenstates
correspond to SCMs {[}Fig.\ \ref{fig3:induceSOSE}(b){]}. Their number
is proportional to $L$, as indicated by the corresponding $w_{\text{sc}}/L$
in Fig.\ \ref{fig3:induceSOSE}(c) and the scaling exponent $\alpha$
in Fig.\ \ref{fig1:main-result}(c) at finite $B$. From Fig.\ \ref{fig3:induceSOSE}(d)
{[}also Fig.\ \ref{fig1:main-result}(d){]},  it is clear that $w_{\text{sc}}/L$
increases monotonically from $0$ at $B=0$ to significant values
($w_{\text{sc}}/L\gtrsim0.01$) at $Ba^{2}\simeq0.1\pi$ for any $|\gamma|\geq|\lambda|$.
These observations demonstrate that SOSE can be induced by magnetic
fields.

It is noteworthy that the field-induced SOSE is always associated
with the formation of non-Hermitian Hofstadter bands with nonzero
Chern numbers in the bulk spectrum {[}Fig.\ \ref{fig3:induceSOSE}(b),
inset{]}. This indicates that the induced SOSE can also be understood
from the emergence of topological line gaps: the chiral-like edge
modes in the line gaps evolve into SCMs by the local non-reciprocity
under OBC.

\begin{figure}[t]
\includegraphics[width=1\linewidth]{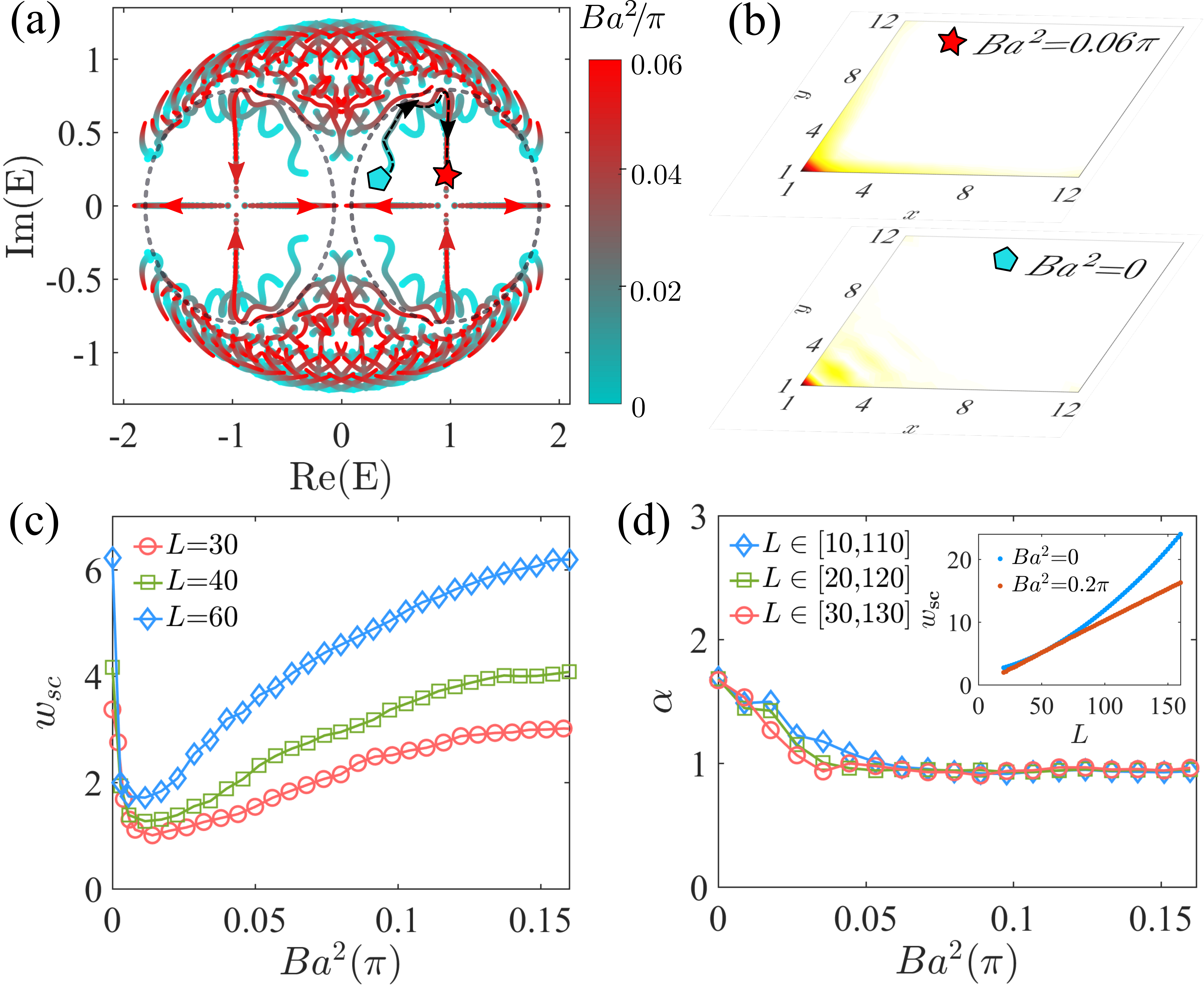}

\caption{(a) Flow of the OBC spectrum of Eq.~\eqref{eq:Hamiltonian2}
as $Ba^{2}$ increases from 0 (cyan) to 0.06$\pi$ (red). (b) $|\psi^{R}({\bf r})|^{2}$
of a SCM at $Ba^{2}=0$ and $0.06\pi$, respectively. (c) $w_{\text{sc}}$
as a function of $Ba^{2}$ for $L=30$, $40$ and $60$, respectively.
(d) $\alpha$ as a function of $Ba^{2}$, extracted by fitting $w_{sc}(L)\text{\ensuremath{\propto}}L^{\alpha}$
with $L\in[10,100]$, $[20,110]$ and $[30,130]$, respectively. Inset:
$w_{\text{sc}}$ as a function of $L$ at $Ba^{2}=0$ and $0.2\pi$,
respectively. $L=12$ in (a,b) and other parameters are $t=\chi=1$,
$h_{x}=h_{y}=0.1,$ and $\xi=1$.}

\label{fig4:pointgap}
\end{figure}

\textit{\textcolor{blue}{Transition from hybrid skin effect to pure
SOSE.}---}Finally, we show that a hybrid skin effect
can emerge in non-Hermitian systems with point-gap topology, and the
system transforms from the hybrid skin effect to pure SOSE by applying
magnetic fields. To be specific, we consider the model\ \cite{Okugawa20prb,Okugawa21prb}
\begin{eqnarray}
\mathcal{H}({\bf k}) & = & M({\bf k})\tau_{0}+i\chi(\sin k_{y}\tau_{x}+\sin k_{x}\tau_{y})-{\bf h}\cdot\bm{\tau},\label{eq:Hamiltonian2}
\end{eqnarray}
where $M({\bf k})=-t(\cos k_{x}+\cos k_{y})$, ${\bf h}=(h_{x},h_{y}),$
and ${\bf \bm{\tau}}=(\tau_{x},\tau_{y})$. $t$, $\chi$ and ${\bf h}$
are real parameters. At zero field, the PBC spectrum has point-gap
topology and encloses a finite area\ \cite{Li2023SM}. Previous
studies identified SOSE in this model, consistent with the point-gap
topology\ \cite{Okugawa20prb,Okugawa21prb}. Remarkably, we find
that FOSE appears simultaneously, leading to a hybrid skin effect.
SOSE generates $\mathcal{O}(L)$ corner modes in the energy regimes
associated with intrinsic second-order topology {[}dashed circles
in Fig.\ \ref{fig4:pointgap}(a){]}~\cite{Okugawa20prb}, while
FOSE generates $\mathcal{O}(L^{2})$ corner modes outside those regimes~\cite{Li2023SM}.
The hybrid skin effect is also corroborated by the scaling exponent
$\alpha$, which is significantly larger than $1$ (the value for
pure SOSE)~\cite{Li2022footnote7-alpha}.

In Fig.\ \ref{fig4:pointgap}, we track the evolution of energies and wavefunctions
of the SCMs associated with SOSE, as the magnetic field $B$ increases from zero to
a finite value. Clearly, they stay inside
the original energy regimes while gradually moving towards the real
axis, as illustrated by red arrows in Fig.~\ref{fig4:pointgap}(a).
Notably, they remain well-localized at the corners during this process
{[}Fig.\ \ref{fig4:pointgap}(b){]}. These features again show the magnetic
robustness of the SCMs associated with SOSE.

With increasing \textbf{ $B$},
the system transforms gradually from the hybrid skin effect to pure
SOSE. To demonstrate this, we plot $w_{sc}$ as a function of $B$
in Fig.\ \ref{fig4:pointgap}(c). For small $B$, $w_{sc}$ drops
suddenly to a finite value, indicating the strong suppression
of FOSE\ \cite{Okuma21prl,LuM21prl,ShaoK22prb}. As $B$ further
increases, $w_{sc}$ grows gradually, indicating the enhancement of
SOSE\ \cite{Li2023SM}. In Fig.\ \ref{fig4:pointgap}(d), we plot
$\alpha$ with increasing $B$. Strikingly, $\alpha$ drops to a plateau
at $1$, which is faster for larger $L$. The scaling behavior of
$w_{sc}$ with respect to $L$ changes from a curved line at zero
field to a straight line at strong fields {[}Fig.\ \ref{fig4:pointgap}(d),
inset{]}. These observations strongly support the transition to
pure SOSE. From a bulk perspective, the magnetic field drives the
energy spectrum with only point gaps to one with non-Hermitian Hofstadter
bands separated by line gaps\ \cite{Li2023SM}.

\textit{\textcolor{blue}{Discussion and conclusion.}---}To summarize, we have presented
a comprehensive description of the influence of magnetic fields on
SOSE. We have shown that magnetic fields enhance and induce SOSE in
non-Hermitian systems. We have also revealed a magnetic
field-driven transition from a hybrid skin effect to pure SOSE in
non-Hermitian systems with point-gap topology. Such anomalous magnetic
robustness and enhancement of SOSE stem from the persistence and emergence
of topological line gaps in the bulk energy spectrum.

Our results can be verified by other models endowed
with SOSE, and generalized to the third-order skin effect~\cite{Li2023SM}. Our predictions can be tested in different physical platforms such
as ultracold atoms, electric circuits and acoustic systems~\cite{Palacios21nc,Zou21nc,Liang22prl,ZhangL21nc,Zhangx21nc}.
Notably, SOSE has been realized in electric circuits recently~\cite{Palacios21nc,Zou21nc}.
Desired effective magnetic fields can be obtained by inserting a uniform
synthetic magnetic flux into the circuits~\cite{LiS22op}. In ultracold-atom
systems, NHSE have also been observed~\cite{Liang22prl}. Highly
tunable gauge fields in this platform have been achieved through lattice
shading~\cite{Hauke12prl} and laser-assisted tunneling~\cite{Miyake13prl,Aidelsburger13prl}.

\begin{acknowledgments}We thank J. Budich and T. Bzdu\v{s}ek for
valuable discussion. C.A.L. thanks the University of Zurich for hospitality
at the early stage of this work. This work was supported by the SFB1170
``ToCoTronics'', the Würzburg-Dresden Cluster of Excellence ct.qmat,
EXC2147, Project-id 390858490, and the High Tech Agenda Bayern, the
European Research Council (ERC) under the European Union's Horizon
2020 research and innovation program (ERC-StG-Neupert-757867-PARATOP).
S.B.Z was also supported by the UZH Postdoc Grant.

\end{acknowledgments}


%

\appendix
\numberwithin{equation}{section}\setcounter{figure}{0}\global\long\def\thefigure{S\arabic{figure}}
\global\long\def\thesection{S\arabic{section}}
\global\long\def\thesubsection{\Alph{subsection}}

\begin{widetext}
\begin{center}
\textbf{\large{}Supplemental material for ``Enhancement of Second-Order Non-Hermitian Skin Effect by Magnetic Fields ''}{\large{} }
\par\end{center}{\large \par}

\section{Absence of point-gap topology under symmetry constraints\label{sec: zero winding number}}

In this section, we prove the absence of point-gap topology for the
minimal model in Eq.\ (2) in the main text {[}see also Eq.\ \eqref{eq:minimum}{]}
from symmetry constraints. The model has finite areas in the bulk
energy spectrum and thus may exhibit FOSE. However, the FOSE is geometry-dependent.
It is absent due to vanishing winding numbers when open boundary conditions
(OBC) are imposed in $x$- and $y$-directions, which we show below.
For a 2D non-Hermitian Hamiltonian $H({\bf k})$, the winding number
in $s=x/y$ direction can be defined as
\begin{equation}
w_{s}\equiv\frac{1}{2\pi i}\int_{0}^{2\pi}dk_{s}\partial_{k_{s}}\mathrm{log\,}\mathrm{det}[H({\bf k})].
\end{equation}

Certain symmetry constraints are able to vanish the winding number,
thus precluding the FOSE. Here, we consider the transpose-associated
mirror symmetry which acts on $H({\bf k})$ as
\begin{equation}
M_{s}H^{T}({\bf k})M_{s}^{-1}=H(g_{s}{\bf k}),
\end{equation}
where $M_{s}$ and $T$ indicate the mirror and transpose operations,
respectively. Explicitly, the group operation $g_{s}$ works as $g_{x}{\bf k}=(-k_{x},k_{y})$
and $g_{y}{\bf k}=(k_{x},-k_{y})$. For the minimal model
\begin{alignat}{1}
H({\bf k}) & =\lambda\sin k_{y}\sigma_{1}+(\gamma+\lambda\cos k_{y})\sigma_{2}+\lambda\sin k_{x}\sigma_{z}-i(\gamma+\lambda\cos k_{x}),\label{eq:minimum}
\end{alignat}
the corresponding transpose operations are $M_{x}=\sigma_{1}$ and
$M_{y}=\sigma_{3}$. When $M_{x}$ symmetry is preserved, we find
that
\begin{align}
w_{x} & =\frac{1}{2\pi i}\int_{-\pi}^{\pi}dk_{x}\partial_{k_{x}}\mathrm{log\,}\mathrm{det}[H(k_{x},k_{y})]\nonumber \\
 & =\frac{1}{2\pi i}\int_{-\pi}^{\pi}dk_{x}\partial_{k_{x}}\mathrm{log}\mathrm{\,det}[M_{x}H^{T}(-k_{x},k_{y})M_{x}^{-1}]\nonumber \\
 & =\frac{1}{2\pi i}\int_{-\pi}^{\pi}dk_{x}\partial_{k_{x}}\mathrm{log\,}\mathrm{det}[H(-k_{x},k_{y})]\nonumber \\
 & =\frac{1}{2\pi i}\int_{\pi}^{-\pi}d(-k_{x})\partial_{-k_{x}}\mathrm{log}\mathrm{\,det}[H(k_{x},k_{y})]\nonumber \\
 & =-w_{x}.
\end{align}
Therefore, $w_{x}=0$. Similarly, we find $w_{y}=0$ when $M_{y}$
is present.

\begin{figure}[t]
\includegraphics[width=1\linewidth]{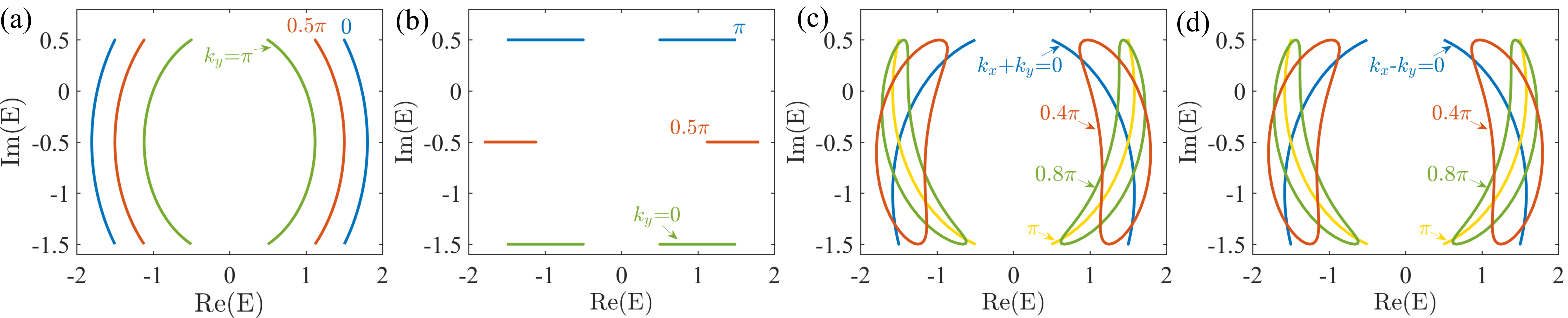}

\caption{(a) Energy spectrum for given $k_{y}$. For any $k_{y}$, the spectrum
consists of two open arcs with no interior in the complex plane. (b)
the same as (a) but for given $k_{x}$. (c) Energy spectrum for given
$k_{x}+k_{y}$. In general, the spectrum for a given $k_{x}+k_{y}$
consists of two loops in the complex plane except for $k_{x}+k_{y}=0$
and $\pi$. (d) the same as (c) but for given $k_{x}-k_{y}$.}

\label{fig:no-point-gap}
\end{figure}

Intuitively, we find that because of the transpose-mirror symmetry,
the states with $(k_{x},k_{y})$ and $(-k_{x},k_{y})$ must be degenerate
in energy. Thus, for any given $k_{x}$ (or $k_{y}$), the spectrum
consists of two open arcs with no interior in the complex plane, as
shown in Figs.\ \ref{fig:no-point-gap}(a) and (b). As a result,
there is no FOSE when OBCs are imposed in $x$- and $y$-directions.
In contrast, for given $k_{x}\pm k_{y}$, we find that in general,
the spectrum consists of two loops with point gaps in the complex
energy plane. This result indicates that the FOSE will appear when
OBC are instead imposed in the diagonal directions (i.e., {[}11{]}
and {[}1$\bar{1}${]} directions).

\section{Non-Hermitian Hofstadter bands\label{sec:Complex-Hofstadter-spectrum}}

In this section, we present the non-Hermitian Hofstadter bands under
strong magnetic fields. To better illustrate the Hofstadter spectrum
in complex energy space, we first derive the Hamiltonian for the minimal
model in Eq.\ \eqref{eq:minimum} in momentum space when the magnetic
flux threading per plaquette is $\phi\equiv Ba^{2}=2\pi p/q$, where
$p$ and $q$ are mutually prime integers. We label the unit cell
as ${\bf R}=j_{x}\hat{e}_{x}+j_{y}\hat{e}_{y}$, where $\hat{e}_{x}$
($\hat{e}_{y}$) is the lattice vector and $j_{x}$ ($j_{y}$) is
an integer. We employ the vector potential ${\bf A}=(0,Bx,0)$ for
the magnetic field $B$. It is convenient to introduce a proper magnetic
unit cell such that the Hamiltonian can be written as
\begin{equation}
H=\int_{0}^{2\pi/q}\frac{dk_{x}}{2\pi/q}\int_{0}^{2\pi}\frac{dk_{y}}{2\pi}{\bf c}^{\dagger}({\bf k})h({\bf k}){\bf c}({\bf k}),
\end{equation}
where ${\bf c}^{\dagger}({\bf k})$ is a $2q$-component vector ${\bf c}^{\dagger}({\bf k})=[c_{1a}^{\dagger}({\bf k}),c_{1b}^{\dagger}({\bf k}),c_{1a}^{\dagger}({\bf k}),c_{2b}^{\dagger}({\bf k}),\cdots,c_{qa}^{\dagger}({\bf k}),c_{qb}^{\dagger}({\bf k})]$;
$h({\bf k})$ is a matrix of $2q$-dimensions. We can rewrite the
Hamiltonian as \citep{Hatsugai06prb}
\begin{equation}
h({\bf k})=\left(\begin{array}{ccccc}
d_{1}({\bf k}) & f^{-} & 0 & \cdots & f^{+}e^{+iqk_{x}}\\
f^{+} & d_{2}({\bf k}) & f^{-} & \cdots & 0\\
0 & f^{+} & \ddots & \ddots & \vdots\\
\vdots & \vdots & \ddots & d_{q-1}({\bf k}) & f^{-}\\
f^{-}e^{-iqk_{x}} & 0 & \cdots & f^{+} & d_{q}({\bf k})
\end{array}\right),\label{eq:BHamiltonian}
\end{equation}
where 
\begin{equation}
d_{j=1:q}({\bf k})=\left(\begin{array}{cc}
-i\gamma & -i\gamma e^{-i\phi j}-i\lambda e^{+ik_{y}}e^{+i\phi j}\\
i\gamma e^{+i\phi j}+i\lambda e^{-ik_{y}}e^{-i\phi j} & -i\gamma
\end{array}\right),
\end{equation}
and
\begin{equation}
f^{+}=\left(\begin{array}{cc}
-i\lambda & 0\\
0 & 0
\end{array}\right),\ \ f^{-}=\left(\begin{array}{cc}
0 & 0\\
0 & -i\lambda
\end{array}\right).
\end{equation}

The magnetic field splits the continuum bulk bands to multiple non-Hermitian
Hofstadter bands with nonzero Chern numbers. To determine the fractal
structure, i.e., the number of Hofstadter bands at specific magnetic
flux $\phi\equiv Ba^{2}=2\pi p/q$ (with $p$ and $q$ being coprime
integers and in units of the flux quantum $h/e$), we need to consider
the real and imaginary parts of the spectrum in the complex plane
simultaneously. Let us focus on the main cases with $p/q=1/N$, with
$N$ being an integer. We find that for even $N$, there are always
$N$ Hofstadter bands. For instance, at $p/q=1/6,$ there are totally
$6$ Hofstadter bands, as shown in Figs.\textcolor{black}{\ \ref{fig:Hofstadter-1}(a)}
and (b). Whereas for odd $N$, the number of Hofstadter bands is either
$2N$ {[}Fig.\ \ref{fig:Hofstadter-1}(c){]} or $N+1$ {[}Fig.\ \ref{fig:Hofstadter-1}(d){]}.
For instance, when $N=5(15)$, there are $10(16)$ Hofstadter bands.
We attribute this particular property to a shifted ``chiral symmetry''
which requires the eigenenergies to come in pairs as $(E+i\gamma,-E+i\gamma)$.
Therefore, the number of Hofstadter bands has to be even.

\begin{figure}[h]
\includegraphics[width=0.6\linewidth]{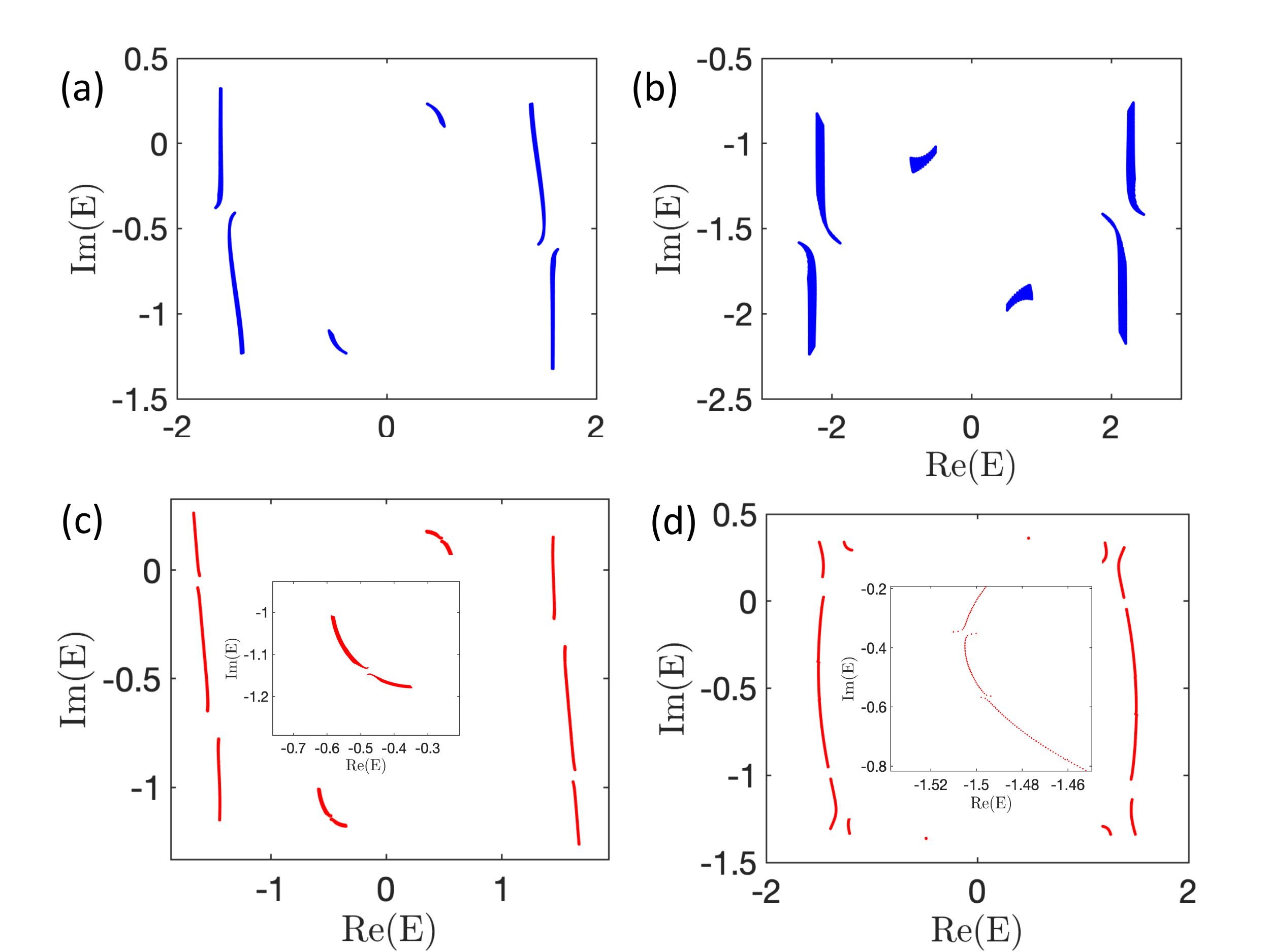}

\caption{Energy spectrum for given magnetic flux $p/q=1/6$ for (a) and (b),
$p/q=1/5$ for (c), and $p/q=1/15$ for (d). Other parameters: $\gamma=0.5$
in (a,c,d), and $\gamma=1.5$ in (b). The flux is in units of the
flux quantum $h/e$. \label{fig:Hofstadter-1}}
\end{figure}

We next show Hofstadter bands developed under strong magnetic fields.\textcolor{blue}{{}
}Figures\ \ref{fig:Hofstadter}(a) and (b) are typical Hofstadter
spectra that display the real and imaginary parts of the bulk spectrum
as functions of rational magnetic flux $\phi$, respectively. Saliently,
we find that fractal structures appear in the real and imaginary Hofstadter
spectra independently. Large line gaps emerge and persist in a broad
range of $\phi$. This feature indicates the emergence of separated
Hofstadter bands and accounts for the robustness and enhancement of
the SOSE under magnetic fields, as discussed above. The number of
Hofstadter bands is determined by the value of rational flux $\phi$
{[}see Fig.\ \ref{fig:Hofstadter}(c){]}, which is similar as discussed
above.

\begin{figure}[t]
\includegraphics[width=0.6\columnwidth]{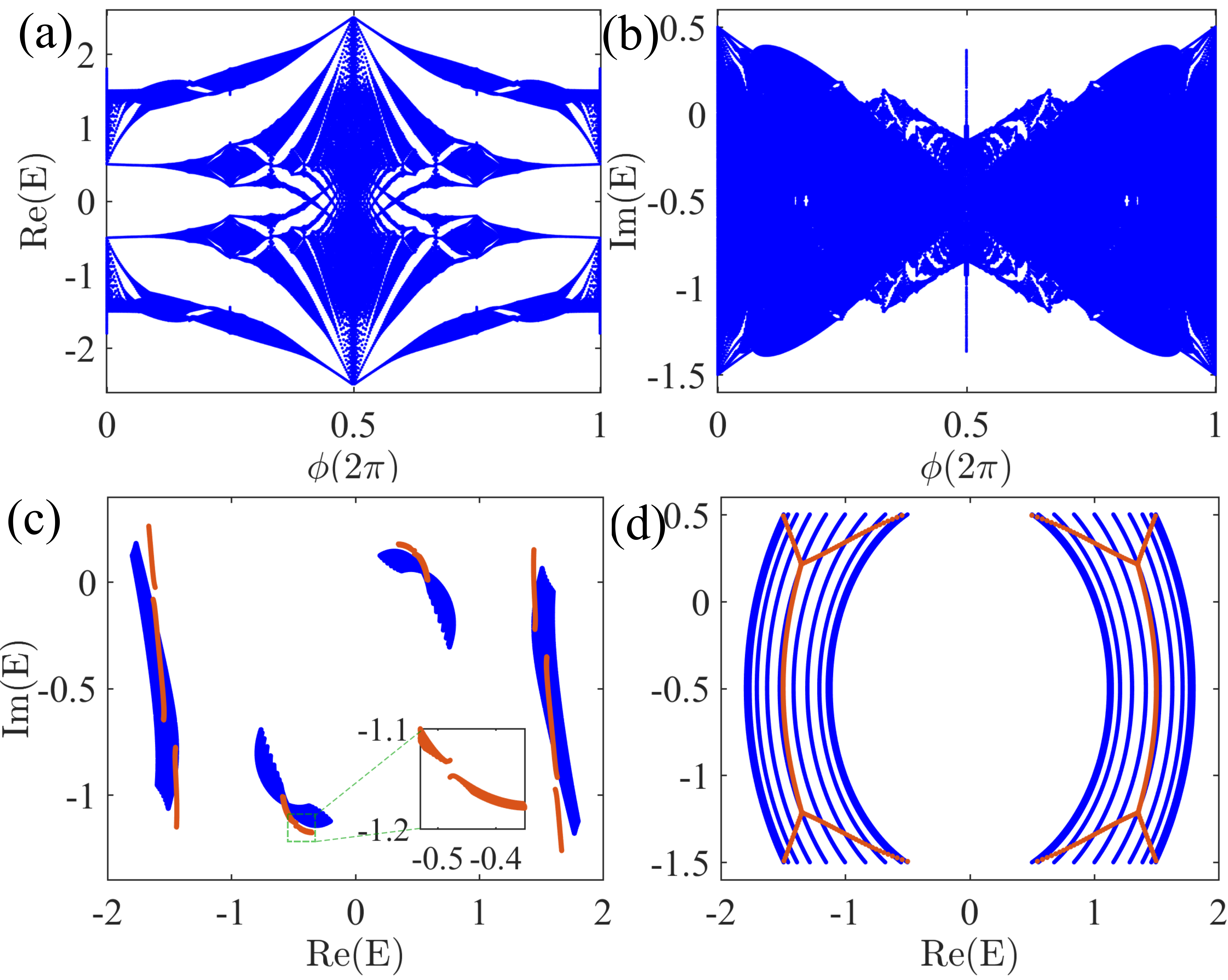}

\caption{(a) and (b): Real and imaginary parts of the Hofstadter spectrum,
respectively. Two extra large bulk gaps appear and persist in the
broad ranges {[}i.e., (0, $\pi$) and $(\pi,2\pi)${]} of $\phi$,
as seen in (a). (c) Energy spectra at $\phi=2\pi/5$ (orange) and
$2\pi/4$ (blue), respectively. (d) Energy spectra at $\phi=0$ (blue)
and $2\pi/400$ (orange), respectively. Other parameters: $\gamma=0.5$,
$\lambda=1$. \label{fig:Hofstadter}}
\end{figure}

Moreover, we observe striking peaks in the Hofstadter spectrum at
some specific flux values. These peaks reflect a high sensitivity
of the bulk spectrum to a small change of the magnetic field. The
spectral sensitivity against magnetic flux grows as the linear system
size $L$ increases. For example, at $B=0$, the bulk spectrum fills
finite areas in the complex plane. When we apply a small field, for
instance, $Ba^{2}=2\pi/400$ to the system, the bulk spectrum collapses
to two line segments maintaining the line gap {[}Fig.\ \ref{fig:Hofstadter}(d){]}.
Note that a finite area in the bulk spectrum indicates the presence
of a geometry-dependent FOSE \citep{Zhangk22nc}. This strong suppression
of spectrum area by small fields is related to the magnetic suppression
of the FOSE, consistent with previous works\citep{Okuma21prl,LuM21prl,ShaoK22prb}.
Similar behavior occurs at other fluxes with small values of $q$.
We emphasize that the drastic change of Hofstadter bands and resulting
anomalous peaks in the Hofstadter spectrum are absent in the Hermitian
limit.

\section{Chern numbers of non-Hermitian Hofstadter bands\label{sec:Non-Hermitian-Chern-number}}

In this section, we calculate the Chern numbers of the non-Hermitian
Hofstadter bands.For a Hofstadter band indexed by $n$, the Chern
number can be calculated as \citep{ShenH18prl,LeeCH19prl,Yao18prl2}
\begin{equation}
C_{n}=\frac{1}{2\pi}\iint dk_{x}dk_{y}[\partial_{k_{x}}A_{k_{y}}^{n}-\partial_{k_{y}}A_{k_{x}}^{n}],
\end{equation}
where the Berry connection is defined as
\begin{equation}
A_{\alpha=k_{x},k_{y}}^{n}\equiv\frac{\langle\psi_{n}^{R}({\bf k})|i\partial_{\alpha}|\psi_{n}^{R}({\bf k})\rangle}{\langle\psi_{n}^{R}({\bf k})|\psi_{n}^{R}({\bf k})\rangle}
\end{equation}
with ${\bf k}=(k_{x},k_{y})$. $|\psi_{n}^{R}({\bf k})\rangle$ satisfies
the eigen function
\begin{equation}
H({\bf k})|\psi_{n}^{R}({\bf k})\rangle=E_{n}({\bf k})|\psi_{n}^{R}({\bf k})\rangle.
\end{equation}
Note that we only employ the right eigenstates. To perform the evaluation
of Chern number effectively, we generalize the method from Ref.\ \citep{Hatsugai06prb}
using the lattice gauge theory, which is proved to be effective for
the Hermitian case.

We first compute the eigen function of the Hamiltonian $h({\bf k})$
in Eq.\ \eqref{eq:BHamiltonian} on meshes in the Brillouin zone
as follows
\begin{equation}
h({\bf k}_{j})|\varphi_{n}^{R}({\bf k}_{j})\rangle=\epsilon_{n}|\varphi_{n}^{R}({\bf k}_{j})\rangle,
\end{equation}
where the momentum points are ${\bf k}_{j}\equiv(j_{x}e_{k_{x}},j_{y}e_{k_{y}})$
with $e_{k_{x}}=\frac{2\pi}{qN_{x}}$ and $e_{k_{y}}=\frac{2\pi}{N_{y}}.$
Here $j_{x}(j_{y})$ is an integer taking values from $0,1,2,\cdots,N_{x(y)}-1$.
We define a $U(1)$ link for each Hofstadter band as \citep{Hatsugai06prb}
\begin{equation}
M_{n,\alpha=x,y}({\bf k}_{j})\equiv|\mathrm{det}U_{n,\alpha}({\bf k}_{j})|^{-1}\mathrm{det}U_{n,\alpha}({\bf k}_{j})
\end{equation}
with the matrix 
\begin{equation}
[U_{n,\alpha}({\bf k}_{j})]_{sp}=\langle\varphi_{n,s}^{R}({\bf k}_{j})|\varphi_{n,p}^{R}({\bf k}_{j}+\hat{e}_{\alpha})\rangle,\ \ (1\leq s,p\leq m),
\end{equation}
where $m$ is the number of states in the $n$-th non-Hermitian Hofstadter
band. This link variables are well-defined except at singular points
with $\mathrm{det}\ U_{\alpha}({\bf k}_{j})=0$. The singularity indicates
the existence of ``vortices'' or ``anti-vortices'' in the wave
function. With this link variable, we obtain a lattice field strength
as
\begin{equation}
F_{n,xy}({\bf k}_{j})\equiv\mathrm{ln}\left[M_{n,x}({\bf k}_{j})M_{n,y}({\bf k}_{j}+\hat{e}_{x})M_{n,x}^{-1}({\bf k}_{j}+\hat{e}_{y})M_{n,y}^{-1}({\bf k}_{j})\right].
\end{equation}
 Finally, the Chern number on the lattice is given by
\begin{equation}
C_{n}=\frac{1}{2\pi i}\sum_{j}F_{n,xy}({\bf k}_{j}).
\end{equation}
This Chern number is proved to be an integer, as it counts the number
of vortices minus the number of anti-vortices. It is numerically verified
to be an integer in the non-Hermitian case considered here.

\section{Evolution of the energy spectrum at small fields\label{sec:bans evolutions}}

In this section, we discuss the spectra of Eq.\ \eqref{eq:minimum}
under small magnetic fields. Figure\ \ref{fig:edgespectra-sensitivity}
presents the evolution of the bulk energy spectrum of a finite-size
system as the magnetic field increases in the small field region.
We find that the area of the bulk spectrum in the complex plane shrinks
rapidly by increasing the magnetic flux. The sensitivity of the bulk
spectrum against magnetic flux becomes more pronounced for a large
system. Notice that a finite area in the bulk spectrum indicates the
presence of the geometry-dependent FOSE \citep{Zhangk22nc}. The strong
suppression of the spectrum area by magnetic fields indeed can reflect
the magnetic suppression of the FOSE.

\begin{figure}[h]
\includegraphics[width=1\linewidth]{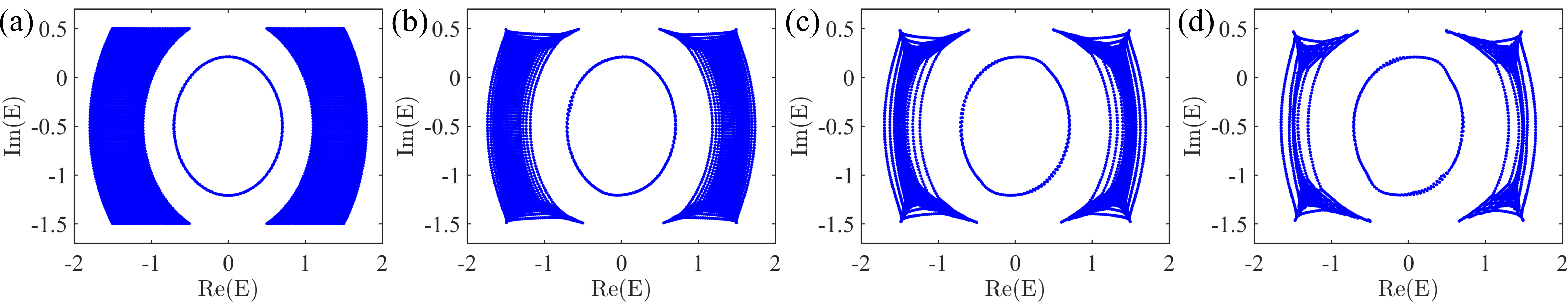}

\caption{Energy spectra on a square geometry under OBC for $L_{x}=L_{y}=100$
and increasing flux densities from $\phi=0$ (a), $6\pi/1000$ (c),
$12\pi/1000$ (c) to $20\pi/1000$ (d) (in units of $h/e$). The area
of the bulk spectrum in the complex energy plane shrinks rapidly by
increasing the flux. \label{fig:edgespectra-sensitivity}}
\end{figure}

\section{Edge spectrum of the models on a ribbon geometry\label{sec:edge spectrum}}

In this section, we demonstrate the existence of 1D edge modes of
the considered models {[}cf. Eqs.\ \eqref{eq:minimum} and \eqref{eq: Hamiltonian2}{]}
on a ribbon geometry. For the minimal model in Eq.\ \eqref{eq:minimum},,
the bulk bands have zero Chern number at zero magnetic field. However,
it exhibits 1D topological modes in a ribbon geometry, as shown in
\textcolor{black}{Fig.\ \ref{fig:edgespectra}(a)}. These edge modes
do not connect to the bulk bands directly but are also protected from
a line-gap topology. They form a closed loop with a point gap in the
complex energy plane. For the model in Eq.\ \eqref{eq: Hamiltonian2},
the bulk bands have nonzero Chern numbers in proper parameter regimes.
Therefore, the chiral edge states connect the bulk bands, as shown
in \textcolor{black}{Fig.\ \ref{fig:edgespectra}(b)}.

\begin{figure}[h]
\includegraphics[width=0.5\linewidth]{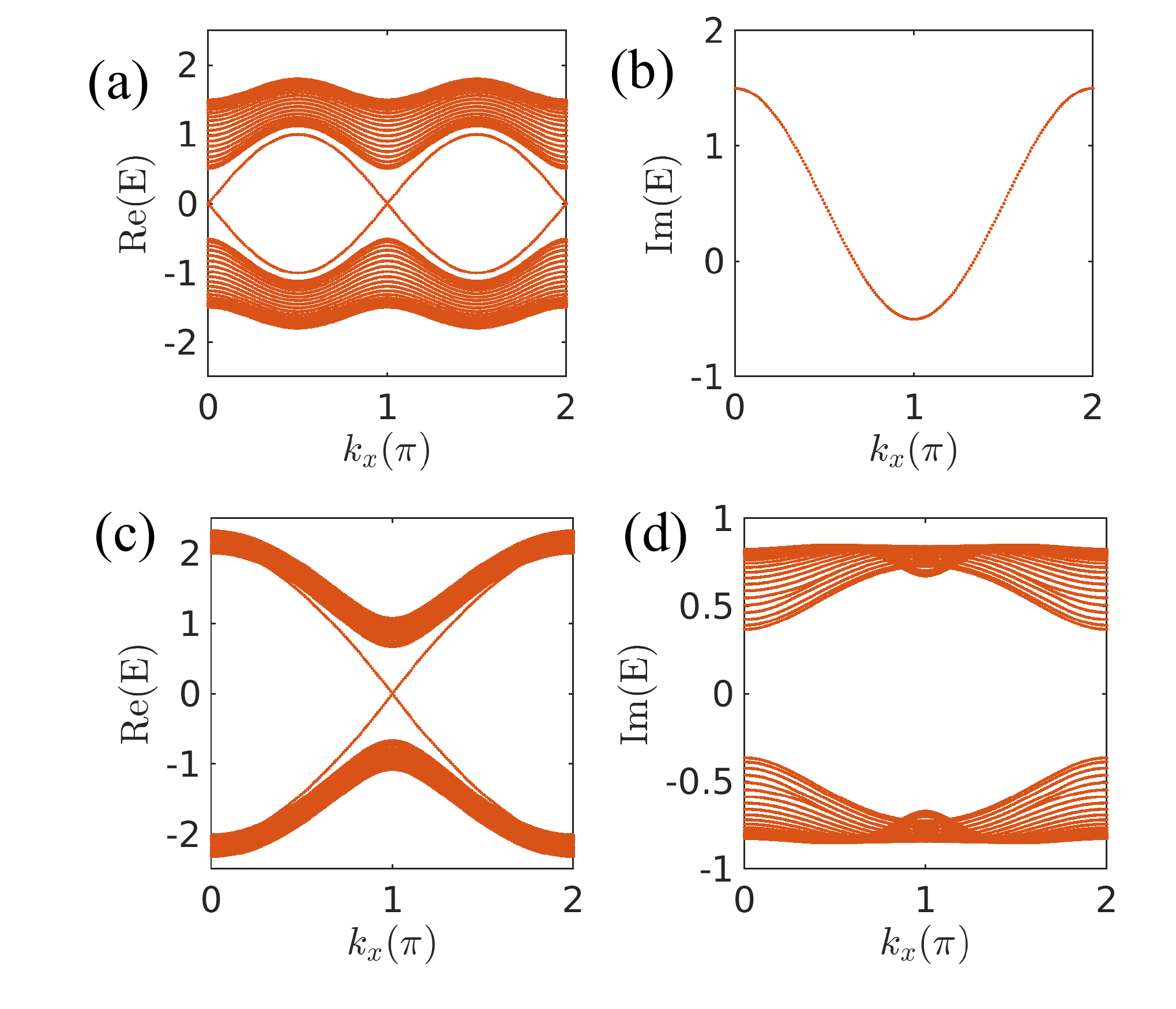}

\caption{(a) and (b) Real and imaginary energy spectra of Eq.\ \eqref{eq:minimum}
on a ribbon geometry with partial OBC in $x$-direction while PBC
in $y$-direction. (c) and (d) are the same as (a) and (b) but for
the model in Eq.\ \eqref{eq: Hamiltonian2}. For (a) and (b), the
parameters are $\gamma=0.5$ and $\lambda=1$. For (c) and (d), parameters
are $\gamma_{x}=t=1,$ $\gamma_{y}=0.2$ and $\delta=0.6$. \label{fig:edgespectra}}
\end{figure}

\section{SOSE model with point gap topology\label{sec:point-gapSOSE}}

In this section, we study the model {[}cf. Eq. (3) in the main text{]}
for the SOSE with point-gap topology. The model has a hybrid skin
effect at zero field, showing both FOSE and SOSE simultaneously. The
OBC spectrum consists of $\mathcal{O}(L)$ modes in the spectral area
of the PBC point gaps, as well as $\mathcal{O}(L^{2})$ \textquotedblleft bulk\textquotedblright{}
modes outside these point gap areas {[}see Fig.\ \ref{fig:spectrumpoint-hybrid}(a)
for the OBC spectrum{]}. The hybrid topology is characterized by the
fact that not only the eigenenergy modes in the nontrivial energy
regions {[}within the dashed circles in Fig.\ \ref{fig:spectrumpoint-hybrid}(a){]},
but also extensive continuum ones (outside the dashed circles) tend
to localize at the corners of the systems under OBC. We confirm this
numerically in Fig.\ \ref{fig:spectrumpoint-hybrid}(b)-(d).

In Fig.\ \ref{fig:spectrumpoint-wsc}(a), we calculate the scaling
exponent $\alpha=d\log(w_{sc})/d\log(L)$ for different $L$ and $\xi$.
We find that $\alpha$ is always significantly larger than $1$ (the
value for the pure SOSE) and it increases as $L$ grows. This further
supports the hybrid skin effect in the system. In Fig.\ \ref{fig:spectrumpoint-wsc}(b),
we calculate the skin corner weight $w_{SC}$ as a function of $L$
for increasing magnetic field. We find that the scaling behavior of
$w_{sc}$ with respect to $L$ takes a curved line at zero field,
while it becomes a nearly straight line at strong fields (i.e., $Ba^{2}>0.05\pi$).

Finally, we present the spectra of the model under magnetic fields.
Without magnetic fields, the model has skin corner modes under OBC
{[}Fig.\ \ref{fig:spectrumpointg}(a){]}. When considering strong
magnetic effects, the PBC spectrum splits to multiple non-Hermitian
Hofstadter bands {[}Fig.\ \ref{fig:spectrumpointg}(b){]}. These
bands carry non-zero Chern numbers, which can be calculated similarly
as that in the previous section (Sec.\ \eqref{sec:Non-Hermitian-Chern-number}).
Saliently, we also find that fractal structures appear in both the
real and imaginary Hofstadter spectra {[}Figs.\ \ref{fig:spectrumpointg}(c)
and \ref{fig:spectrumpointg}(d){]}. At small fields, it shows the
formation of Landau levels. It is also worthy noting that line gaps
quickly develop when applying a magnetic field, as shown in Fig.\ \ref{fig:spectrumpointg}(d).

\begin{figure}[h]
\includegraphics[width=0.6\linewidth]{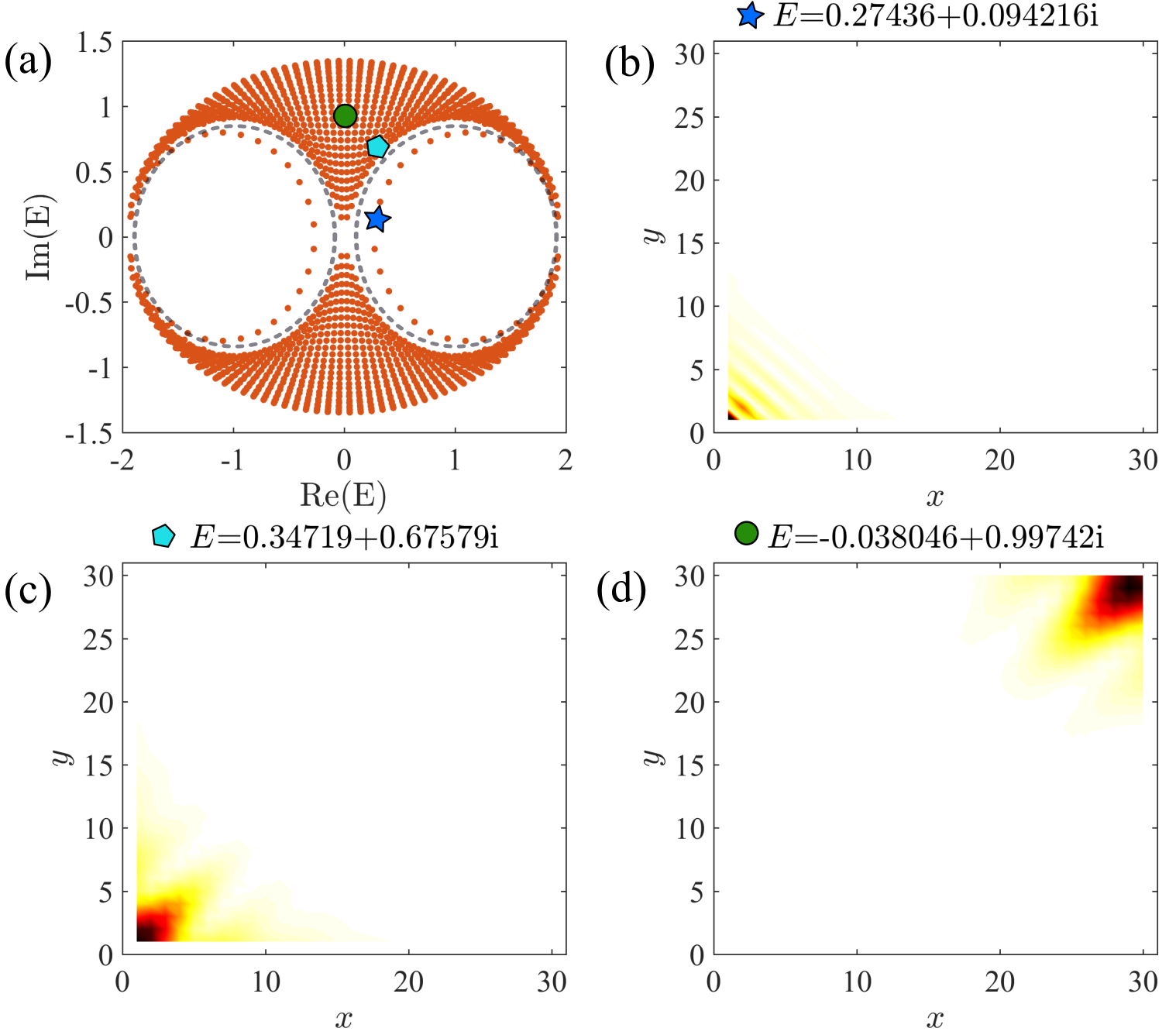}

\caption{Energy spectrum under full OBC at zero field. The modes with energies
inside the dashed circles resemble the skin corner modes associated
with intrinsic second-order topology (point gaps in the bulk spectrum).
(b) Wavefunction distribution $|\psi^{R}|{}^{2}$ of a mode with energy
inside the circles {[}indicated by the dark blue star in (a){]}. (c)
and (d) are the same as (b) but for two modes outside the circles
(indicated by the cyan pentagon and green disk, respectively). Other
parameters are the same as Fig.\ 4 in the main text. \label{fig:spectrumpoint-hybrid}}
\end{figure}

\begin{figure}[th]
\includegraphics[height=5cm]{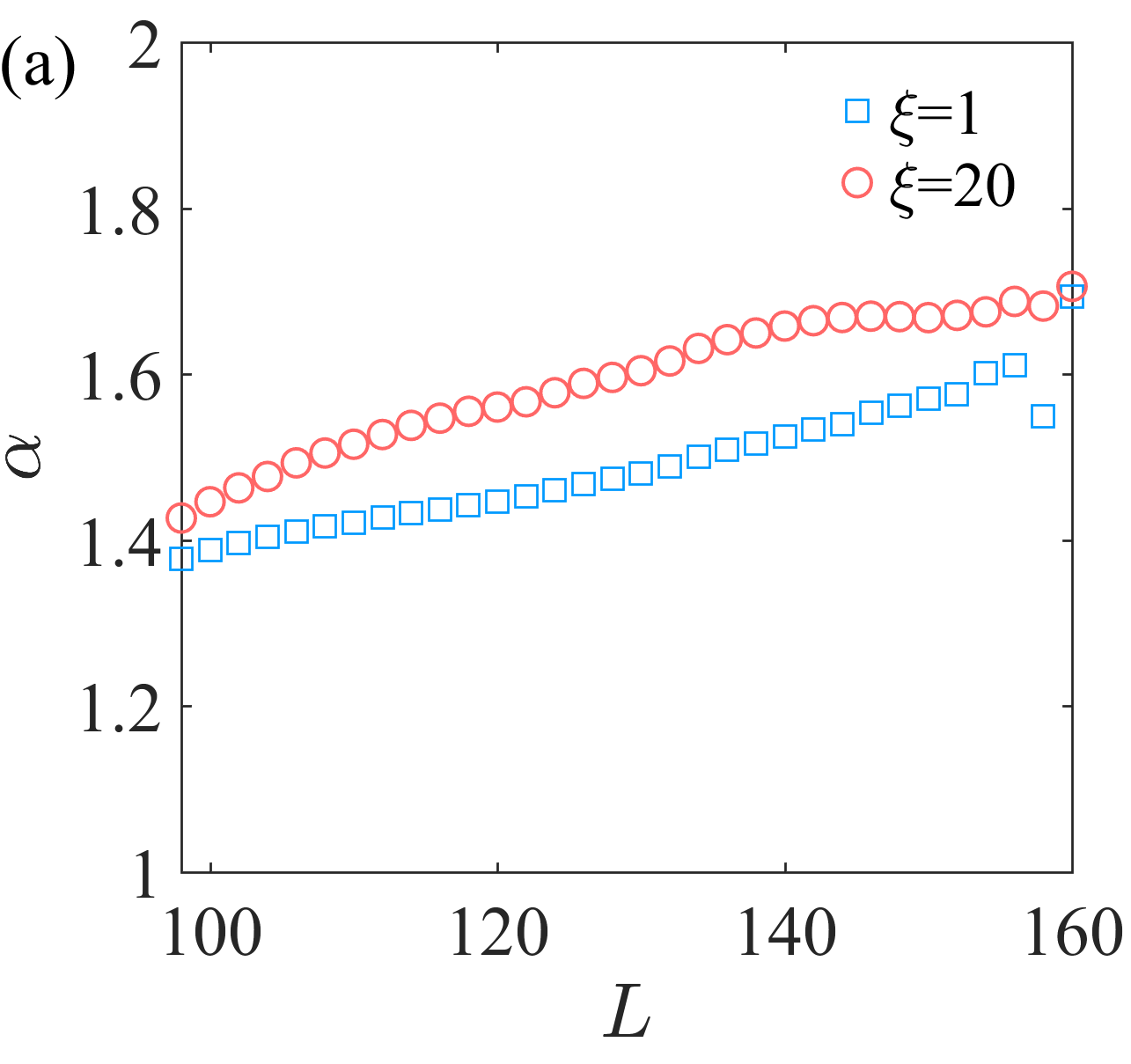}\ \ \ \ \ \ \includegraphics[height=5cm]{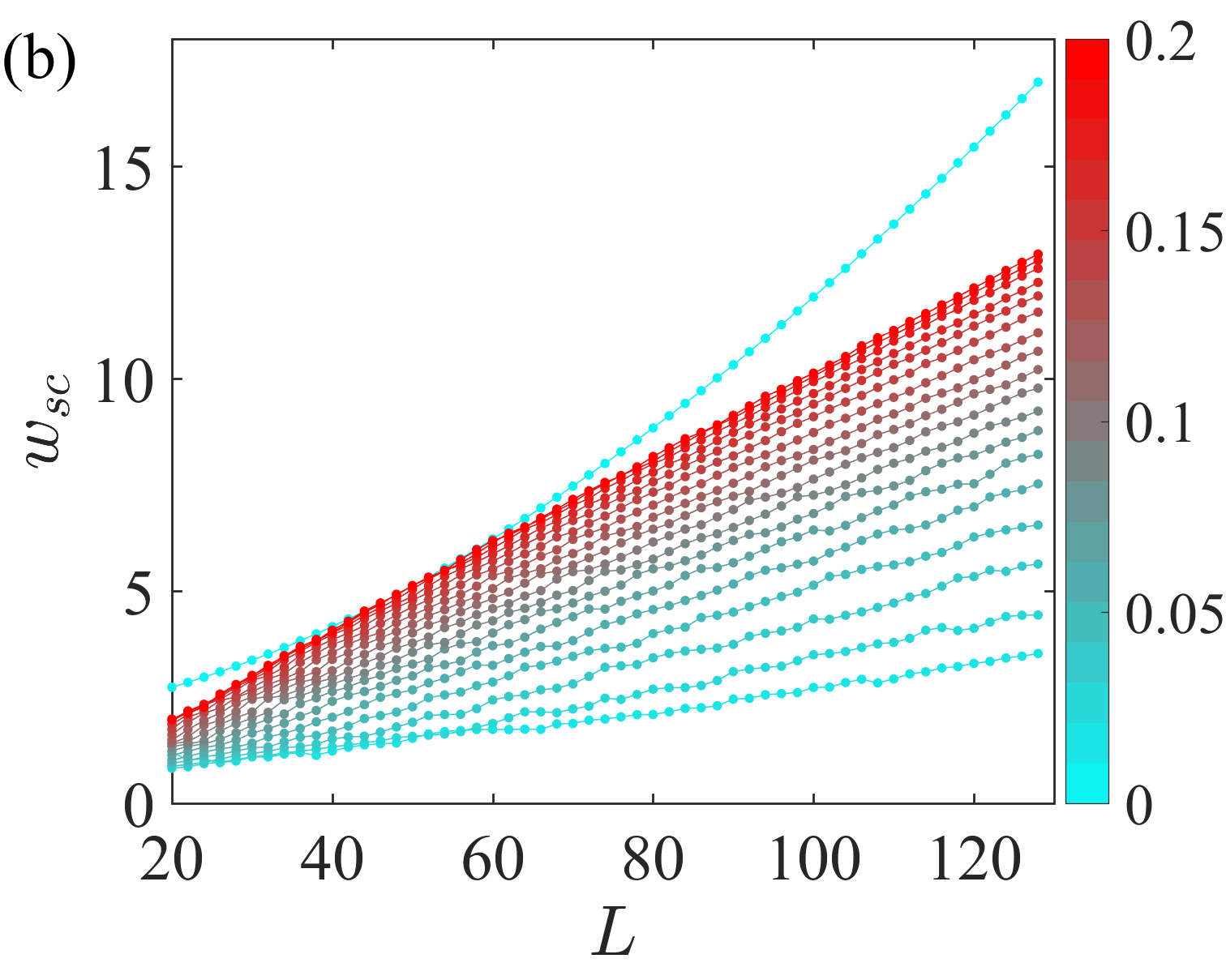}

\caption{(a) Scaling exponent $\alpha=d\log(w_{sc})/d\log(L)$ as a function
of system size $L$. (b) Skin corner weight $w_{sc}$ as a function
of $L$ for increasing $Ba^{2}$ from 0 (cyan) to 0.02$\pi$ (red).
Parameters: $\xi=1$ in (b), and other parameters are the same as
Fig.\ 4 in the main text. \label{fig:spectrumpoint-wsc}}
\end{figure}

\begin{figure}[th]
\includegraphics[width=0.8\linewidth]{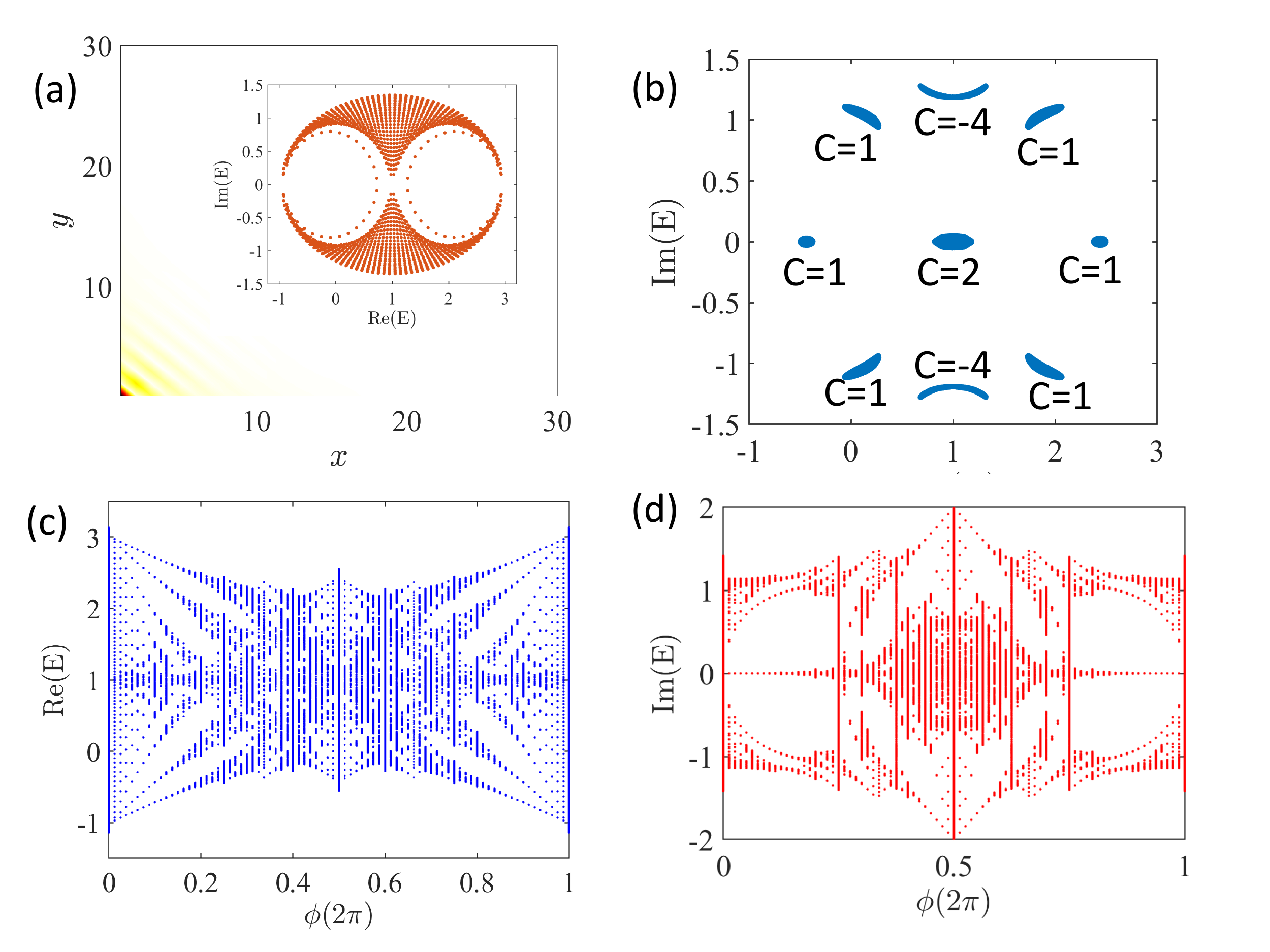}

\caption{(a) Energy spectrum at zero field (inset) under OBC and distribution
$|\psi^{R}({\bf r})|^{2}$ of a skin corner mode with energy $E=1.27+0.09i$.
(b) Non-Hermitian Hofstadter bands at $Ba^{2}=0.4\pi$ under PBC.
The numbers $C$ stand for the Chern numbers of the bands. Real (c)
and imaginary (d) parts of energy spectrum under PBC as a function
of magnetic flux $\phi=Ba^{2}$. Other parameters are the same as
Fig.\ 4 in the main text. \label{fig:spectrumpointg}}
\end{figure}

\section{Generalization to other models with SOSE\label{sec:generalization to other models}}

In this section, we generalize our result to another model of the
SOSE with line gap. We have shown that the SOSE is robust and can
even be enhanced under magnetic fields. To better support the generality
and experimental feasibility of our theory, we further consider a
different model with the SOSE \citep{LeeCH19prl,Zhuw22prb}
\begin{alignat}{1}
H_{2}({\bf k}) & =(\gamma_{x}+t\cos k_{x})\tau_{1}\sigma_{0}-t\sin k_{x}\tau_{2}\sigma_{3}+i\delta\tau_{1}\sigma_{2}\nonumber \\
 & -(\gamma_{y}+t\cos k_{y})\tau_{2}\sigma_{2}-t\sin k_{y}\tau_{2}\sigma_{1}-i\delta\tau_{2}\sigma_{0},\label{eq: Hamiltonian2}
\end{alignat}
where $\sigma_{i}$ and $\tau_{i}$ ($i=1,2,3$) are Pauli matrices
for the four sublattice degrees of freedom in a unit cell. $t$ is
the strength of intercell hopping, and $\gamma_{x/y}\pm\delta$ the
nonreciprocal hopping amplitudes occurring within unit cells in $x/y$-directions.
Note that this model has been realized experimentally \citep{Zou21nc}.
For $|\gamma_{x}|<t$ (or $|\gamma_{y}|<t$) and $|\delta|>0$, the
model shows the SOSE, which can be understood similarly based on edge
modes. The FOSE is prohibited by the presence of two transpose-mirror
symmetries, similar to the minimal model in the main text. However,
the edge modes in this model are chiral-like and appear when OBC are
imposed to either direction, different from the minimal model. For
$|\gamma_{x}|>t$ and $|\gamma_{y}|>t$, the SOSE is absent in the
model at $B=0$ for any $\delta$.

\begin{figure}
\includegraphics[width=0.6\linewidth]{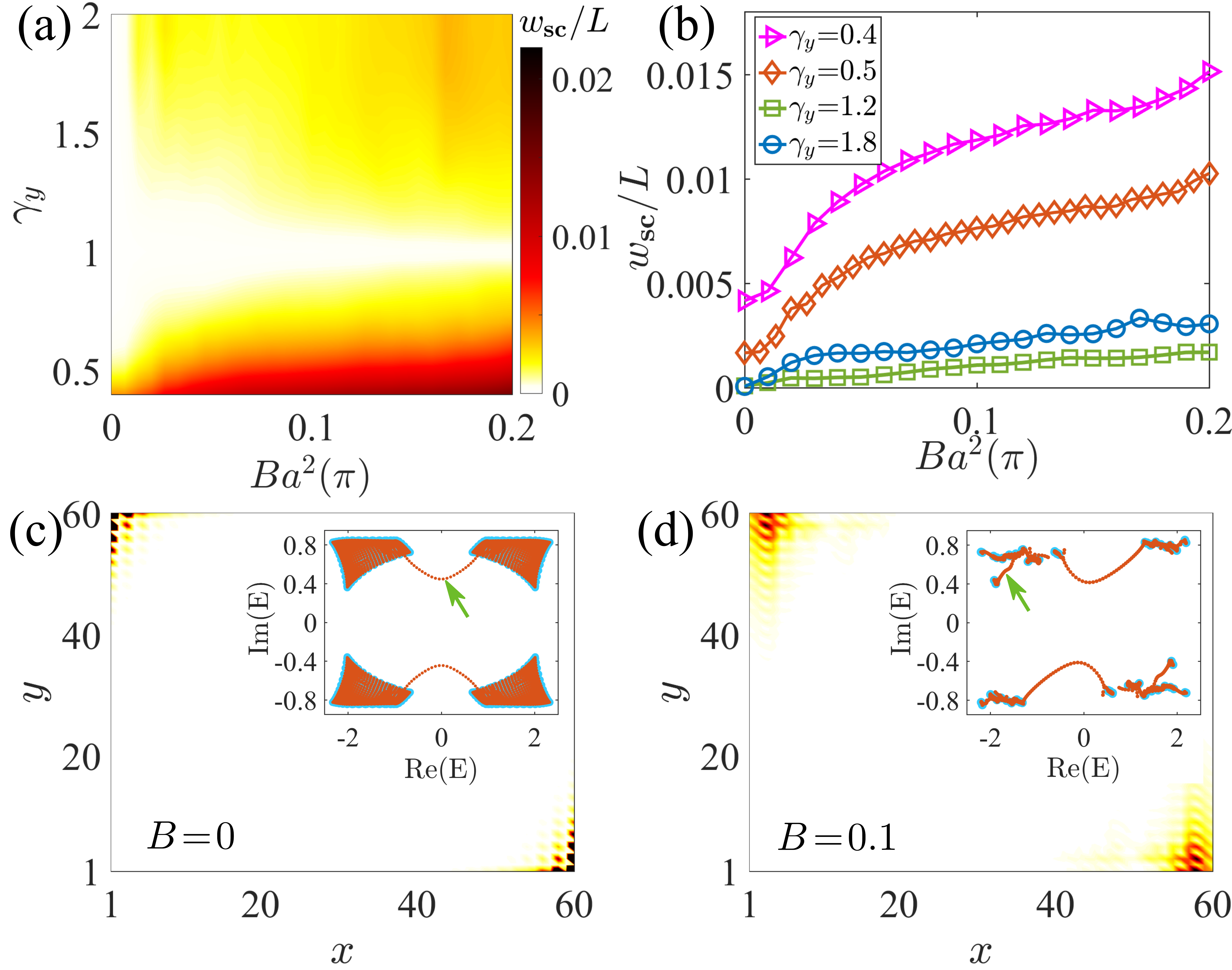}

\caption{(a) $w_{\text{sc}}/L$ as a function of $B$ and $\gamma_{y}$. (b)
$w_{\text{sc}}/L$ as a function of $B$ for different $\gamma_{y}$.
(c) $|\psi^{R}({\bf r})|^{2}$ of a skin corner mode at $B=0$ under
OBC. Inset: the PBC (cyan) and OBC (orange) energy spectra. The arrow
marks the mode energy $E=0.04\text{\ensuremath{-}}0.45i$. (d) the
same as (a) but for $Ba^{2}=0.1\pi$. The arrow marks the mode energy
$E=-1.73\text{\ensuremath{+}}0.49i$. Other parameters: $\gamma_{x}=t=1,$
$\gamma_{y}=0.2$, $\delta=0.6$ and $L=60$. \label{fig4:Lee}}
\end{figure}

\begin{figure}
\includegraphics[width=0.6\linewidth]{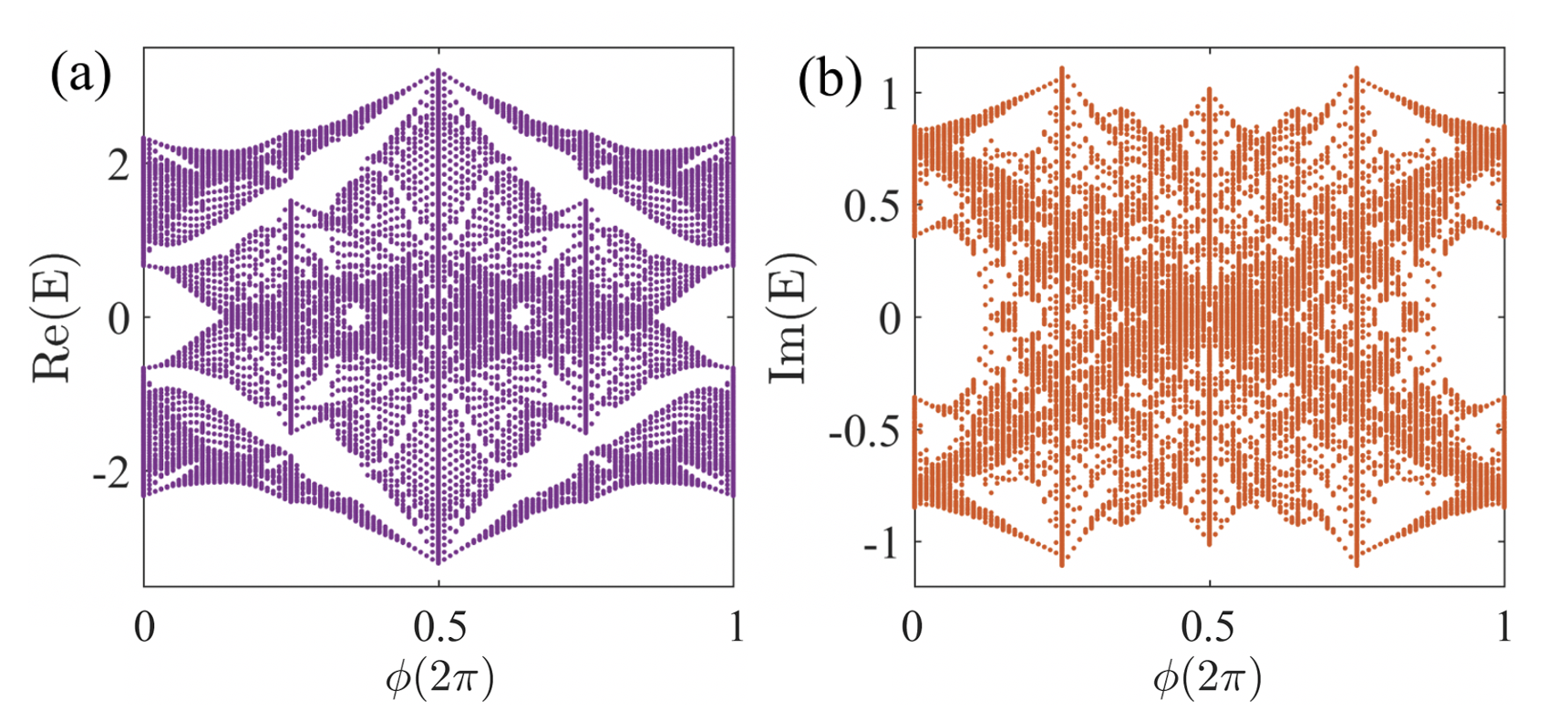}

\caption{Real (a) and imaginary (b) parts of energy spectrum as a function
of magnetic flux $\phi=Ba^{2}$ under PBC. Other parameters: $\gamma_{x}=t=1,$
$\gamma_{y}=0.2$, $\delta=0.6$. \label{fig4:Lee-butterfly}}
\end{figure}

Using the formula of skin corner weight defined in the main text,
we calculate $w_{\text{sc}}$ of the model in Eq.\ \eqref{eq: Hamiltonian2}
under OBC as increasing \textcolor{black}{magnetic field }$B$.\textcolor{black}{{}
For illustration, we consider $\gamma_{x}=t=1$ and $\delta=0.6$
and present $w_{\text{sc}}/L$} as a function of $B$ for different
$\gamma_{y}$ {[}Figs.\ \textcolor{black}{\ref{fig4:Lee}(a) and
(b)}{]}. At $B=0$\textcolor{black}{, $w_{\text{sc}}/L$} is vanishingly
small for large $\gamma_{y}(>t$), whereas it becomes considerable
for small $\gamma_{y}(<t$). This confirms the absence (presence)
of the SOSE for $\gamma_{y}>t$ ($\gamma_{y}<t$). Evidently, $w_{\text{sc}}$
increases significantly under $B$ for proper parameters. These features
clearly demonstrate the enhancement (for $\gamma_{y}<t$) and emergence
(for $\gamma_{y}>t$) of SOSE by \textcolor{black}{magnetic fields.
We can also find that the SOSE is ultimately related to the formation
of topological line gaps in the bulk spectrum under $B$, as shown
in }Fig.\ \textcolor{black}{\ref{fig4:Lee}(d).}

We further plot the real and imaginary parts of the spectrum of the
model in Eq.\ \eqref{eq: Hamiltonian2} as a function of magnetic
flux in Figs.\ \textcolor{black}{\ref{fig4:Lee-butterfly}(a) and
(b), respectively. Both spectra exhibit Hofstadter\textquoteright s
fractal-like structures independently. We argue here that the major
difference lies in rich internal degrees of freedom in our considered
}model Eq.\ \eqref{eq: Hamiltonian2}\textcolor{black}{. We also
note the persistence and emergence of topological line gaps, which
accounts for the anomalous magnetic robust behaviors discussed above.}

\section{Magnetic suppression of the first-order skin effect\label{sec:Results-for-the}}

In this section, we show the magnetic suppression of the FOSE. The
FOSE may also exhibit skin corner modes. We demonstrate here that
the magnetic field will significantly suppress the skin corner modes
arising from the first-order nature, rather than showing magnetic
robustness or enhancement. To illustrate this point explicitly, we
consider the following model \citep{LiuT19prl}

\begin{alignat}{1}
H_{\text{1st}}({\bf k}) & =(\gamma{}_{x}+\lambda\cos k_{x})\tau_{1}\sigma_{0}-(\lambda\sin k_{x}+i\delta)\tau_{2}\sigma_{3}\nonumber \\
 & -(\gamma{}_{y}+\lambda\cos k_{y})\tau_{2}\sigma_{2}-(\lambda\sin k_{y}+i\delta)\tau_{2}\sigma_{1},\label{eq: Hamiltonian2-1}
\end{alignat}
where $\sigma_{i}$ and $\tau_{i}$ ($i=1,2,3$) are Pauli matrices
for the four sublattice degrees of freedom in a unit cell. The nonreciprocal
hopping occurs in both $x$- and $y$-directions with the corresponding
hopping amplitudes $\gamma_{x/y}\pm\delta$. This model exhibits skin
corner modes but whose number scales with the area of the system $L^{2}$
in 2D, which is of FOSE by definition. In contrast to the models of
the SOSE, the nonreciprocal hoppings in Eq.~\eqref{eq: Hamiltonian2-1}
are finite throughout the system. An effective nonreciprocal vector
in the bulk can be found, which points in {[}11{]} direction. Note
that the model reduces to the Benalcazar-Bernevig-Hughes model in
the Hermitian limit of $\delta=0$ \citep{BBH17prb,Benalcazar17Science}.

As shown in \textcolor{black}{Fig.\ \ref{fig:1storder}(a)}, the
energy spectrum of the model under OBC is dramatically different from
that under full PBC: the PBC spectrum (for $L\rightarrow\infty$)
covers a finite area with point-gap topology in the complex-energy
plane, while the OBC spectrum forms two separated lines on the real
axis (under chosen parameters). Under OBC, we observe that all eigenstates
of the system are localized to the left-bottom corner, forming skin
corner modes. When applying a magnetic field, we find that an extensive
number of corner states are pushed away from corners due to the strong
magnetic confinement effect and the breakdown of point-gap protection
\citep{LuM21prl}\textcolor{black}{{} {[}see Fig.\ \ref{fig:1storder}(b)
for an illustration{]}.} We again employ the previously defined skin
corner weight $w_{sc}$ to characterize the magnetic response in present
case. Such a suppression of the first-order skin corner modes is clearly
indicated by the obvious decrease of $w_{sc}$ as increasing magnetic
field \textcolor{black}{strength {[}Fig.\ \ref{fig:1storder}(c){]}.
This result clearly distinguishes the magnetic response of the FOSE
from the second-order one and reveals their different topological
properties.}

\begin{figure}[t]
\includegraphics[width=0.9\linewidth]{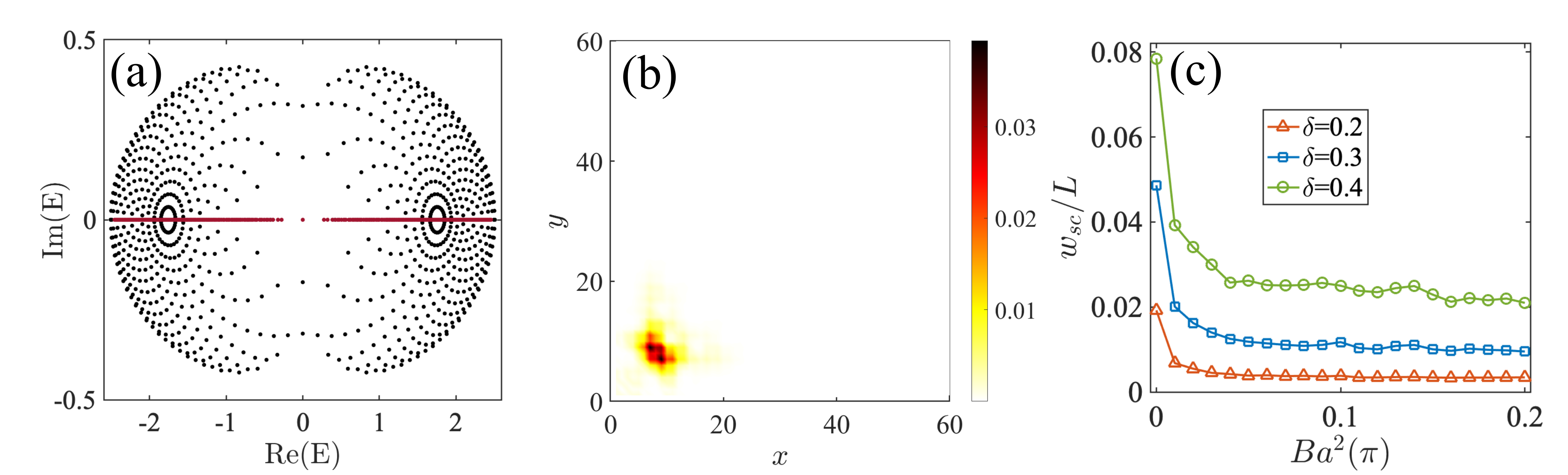}

\caption{Magnetic suppression of the first-order skin corner modes. (a) Energy
spectra for the PBC (black) and the OBC (red), respectively. (b) Wavefunction
$|\psi^{R}({\bf r})|^{2}$ of an eigenenergy state with energy $E=-0.6$
at $Ba^{2}=\pi/10$. The state is moved away from the corner by the
magnetic field $B$. (c) $w_{sc}$ as a function of $B$ for different
nonreciprocal strengths $\delta$. Other parameters are $\gamma{}_{x}=\gamma{}_{y}=0.8$,
$\lambda=1$ and $L=60$ in all panels, and $\delta=0.3$ in (a) and
(b).}

\label{fig:1storder}
\end{figure}

\section{Magnetic robustness of the third-order skin effect \label{sec:3rd order SE}}

In this section, we discuss the magnetic response of the third-order
skin effect. To be specific, we consider the prototypical model\ \citep{Kawabata20prb}
\begin{equation}
H_{\mathrm{3rd}}=i\lambda\sin k_{y}\sigma_{x}+i(\gamma+\lambda\cos k_{y})\sigma_{y}+i\lambda\sin k_{x}\sigma_{z}+(\gamma+\lambda\cos k_{x})\tau_{z}+\lambda\sin k_{z}\tau_{y}+(\gamma+\lambda\cos k_{z})\tau_{x},\label{eq:3rd-order}
\end{equation}
where $\bm{\sigma}$ and $\bm{\tau}$ are Pauli matrices in sublattice
spaces, $\gamma$ and $\lambda$ are real parameters. Under OBC, there
are $\mathcal{O}(L)$ skin corner modes emerge in all $\mathcal{O}(L^{3})$
modes, as shown in Fig.\ \ref{fig:3rd}(a). When applying a magnetic
field, we find that the skin corner modes survive in the line gap
{[}Fig.\ \ref{fig:3rd}(b){]}. Similarly, we define the skin corner
weight as 
\begin{equation}
w_{sc}=\sum_{n,{\bf r},{\bf r}_{c}}|\psi_{n}^{R}({\bf r})|^{4}\exp(-|{\bf r}-{\bf r}_{c}|/\xi),
\end{equation}
similar to the 2D case, in which all eight corners are taken into
account. Figure\ \ref{fig:3rd}(c) plots $w_{sc}$ as a function
of magnetic field $B$. We see that $w_{sc}$ increases significantly
as $B$ increases. Thus, we conclude the magnetic robustness and enhancement
of the third-order skin effect. The underlying mechanism is similar
to that of the SOSE with line gaps.

\begin{figure}
\includegraphics[width=0.9\linewidth]{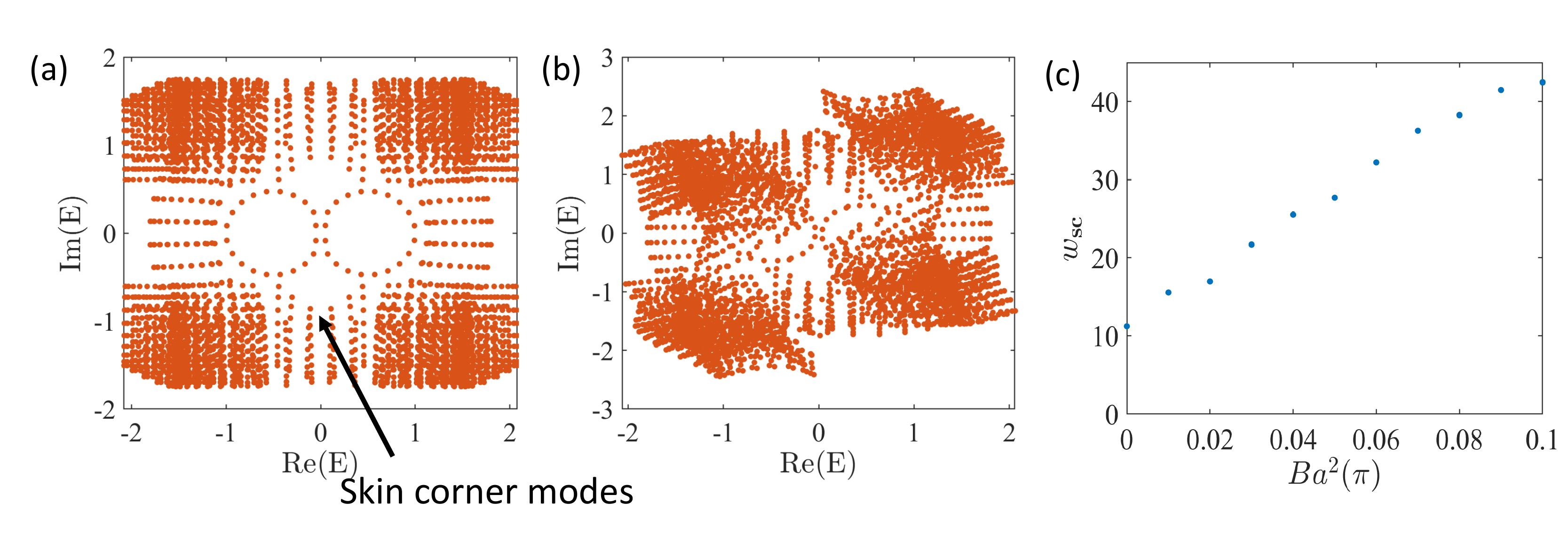}

\caption{OBC energy spectrum of the model in Eq.\ \eqref{eq:3rd-order} with
the third-order skin effect at $Ba^{2}=0$ (a) and $0.04\pi$ (b).
(c) Skin corner weight as a function of $Ba^{2}$. Other parameters:
$\lambda=1,$ $\gamma=0.5$, $L=10$. \label{fig:3rd}}
\end{figure}

\end{widetext}


\begin{thebibliography}{102}
\bibitem [{Li2({\natexlab{a}})}]{Li2022footnote1-FOSE}%
  \BibitemOpen
  \href@noop {} {\bibfield  {journal} {\bibinfo  {journal} {When the bulk
  spectrum covers finite areas in the complex plane, FOSE may be
  geometry-dependent~\cite{Zhangk22nc} and appear when OBC are imposed to a
  particular direction. These symmetries can enforce degeneracy of
  eigenenergies, preventing point-gap topology in the complex bulk energy
  spectrum in specific directions. Thus, FOSE is forbidden when OBC are applied
  to the corresponding directions}\ }}\BibitemShut {NoStop}%
\bibitem [{Li2({\natexlab{d}})}]{Li2022footnote3-FOSE}%
  \BibitemOpen
  \href@noop {} {\bibfield  {journal} {\bibinfo  {journal} {FOSE (if
  present) also features skin corner modes and contribute to $w_{\text{sc}}$
  when the system boundaries are not perfectly aligned with the direction of
  the non-reciprocal hopping. However, in this case, $w_{\text{sc}}$ scales
  quadratically with $L$, as we have verified in the SM}\ }}\BibitemShut {NoStop}%
\bibitem [{Li2({\natexlab{f}})}]{Li2022footnote4-alpha}%
  \BibitemOpen
  \href@noop {} {\bibfield  {journal} {\bibinfo  {journal} {Note that
  $\log(w_{\text{sc}})/\log(L)$ becomes equivalent to the scaling exponent
  $\alpha$ when $L$ and $\xi$ are much larger than the localization length of
  the SCMs}\ }}\BibitemShut {NoStop}%
\end{thebibliography}
\end{document}